\documentclass[twocolumn, showkeys, superscriptaddress]{revtex4-2}
\usepackage[T2A]{fontenc}
\usepackage[utf8]{inputenc}
\usepackage[american]{babel}
\usepackage{array}
\usepackage{units}
\usepackage{textcomp}
\usepackage{multirow}
\usepackage{amsmath}
\usepackage{amssymb}

\usepackage{graphicx}

\providecommand{\tabularnewline}{\\}

\usepackage{fancyhdr}
\fancypagestyle{firstpage}{%
\lhead{\footnotesize{This is the pre-peer reviewed version of the following article: Voronenkov, V., Groven, B., Medina Silva, H., Morin, P. and De Gendt, S. (2024), Guiding Principles for the Design of a Chemical Vapor Deposition Process for Highly Crystalline Transition Metal Dichalcogenides. Phys. Status Solidi A 2300943., which has been published in final form at \href{https://doi.org/10.1002/pssa.202300943}{https://doi.org/10.1002/pssa.202300943}. This article may be used for non-commercial purposes in accordance with Wiley Terms and Conditions for Use of Self-Archived Versions.}}
  \rhead{}
}

\begin{document}

\title{Guiding principles for the design of a chemical vapor deposition process for highly crystalline transition metal dichalcogenides}

\author{Vladislav~Voronenkov}
\affiliation{KULeuven, Faculty of Science, Celestijnenlaan 200F, B-3001 Leuven, Belgium}
\affiliation{imec, Department of Advanced Patterning, Process and Materials, Kapeldreef 75, 3001 Leuven, Belgium}
\author{Benjamin~Groven} \email[Corresponding author: ]{benjamin.groven@imec.be}
\affiliation{imec, Department of Advanced Patterning, Process and Materials, Kapeldreef 75, 3001 Leuven, Belgium}
\author{Henry~Medina~Silva}
\affiliation{imec, Department of Advanced Patterning, Process and Materials, Kapeldreef 75, 3001 Leuven, Belgium}
\author{Pierre~Morin}
\affiliation{imec, Department of Advanced Patterning, Process and Materials, Kapeldreef 75, 3001 Leuven, Belgium}
\author{Stefan~De~Gendt}
\affiliation{KULeuven, Faculty of Science, Celestijnenlaan 200F, B-3001 Leuven, Belgium}
\affiliation{imec, Department of Advanced Patterning, Process and Materials, Kapeldreef 75, 3001 Leuven, Belgium}
\keywords{2D materials, chemical vapor deposition, transition metal dichalcogenides}

\begin{abstract}

  Two-dimensional transition metal dichalcogenides  (TMDs) for advanced logic transistor technologies are deposited by various modifications of the chemical vapor deposition (CVD) method using a wide variety of precursors.
  Being a major electrical performance limiter, the TMD crystal grain size strongly differs between the various CVD precursor chemistries from nano- to millimeter-sized crystals.
However, it remains unclear how the CVD precursor chemistry affects the nucleation density and resulting TMD crystal grain size.
This work postulates guiding principles to design a CVD process for highly crystalline TMD deposition using a quantitative analytical model benchmarked against literature.
The TMD  nucleation density reduces favorably under low supersaturation conditions, where the metal precursor sorption on the starting surface is reversible and the corresponding metal precursor desorption rate exceeds the overall deposition rate.
Such reversible precursor adsorption guarantees efficient long-range gas-phase lateral diffusion of precursor species in addition to short-range surface diffusion, which vitally increases crystal grain size.
As such, the proposed model explains the large spread in experimentally observed TMD nucleation densities and crystal grain sizes for state-of-the-art CVD chemistries.
Ultimately, it empowers the reader to interpret and modulate precursor adsorption and diffusion reactions through designing CVD precursor chemistries compatible with temperature sensitive application schemes.

\end{abstract}

\maketitle
\thispagestyle{firstpage}

\section{Introduction}

\subsection{Motivation}

Semiconducting two-dimensional (2D) transition metal dichalcogenides (TMDs) with the generic formula MX\textsubscript{2}, where M is a transition metal and X is a chalcogen, promise to complement silicon (Si) as a channel material by providing better carrier mobility in densely packed ultra-scaled digital circuits \cite{robinson2018perspective,wang2021road}.
In particular, molybdenum and tungsten disulfide (MoS$_2$ and WS$_2$) have attracted particular research interest due to the combination of sufficiently high carrier mobility \cite{schmidt2015electronic,li2016charge,manzeli20172d} and relative stability in the ambient environment \cite{yang2022oxidations}.
In recent years, the field has advanced from single transistors based on mechanically exfoliated flakes of MoS$_2$ and WS$_2$ \cite{Radisavljevic2011,Georgiou2013},
to wafer-scale fabrication of logic gates and SRAM \cite{li2020large}. 

However, many challenges remain before TMD materials can be integrated into sophisticated devices. One of these is the deposition of a  high quality TMD monolayer (ML) on 300~mm wafers, which is essential for the envisioned industrial applications.
Bulk crystals of MoS$_2$ and WS$_2$ cannot be grown from the melt as these materials decompose at temperatures well below the melting point. In the absence of a native single-crystal substrate, MoS$_2$ and WS$_2$ polycrystalline layers are typically deposited from vapor phase on non-native crystalline or amorphous substrate.

Crystal grain boundaries in TMD material remain a key intrinsic material defect type limiting the TMD electrical performance in semiconductor devices \cite{Ly2016grain-boundary-scattering,Kim2017grain-boundary-mobility}.
Such intergranular boundaries are formed by the coalescence of individual crystalline TMD nuclei during the deposition process. Thus, it is possible to increase the grain size and proportionally decrease the density of intergrain defects by decreasing the TMD nucleation density.
Therefore, one of the key requirements for a TMD deposition process is the ability to control the nucleation density in order to reduce the density of defects that limit the electrical device performance. 

Here, the TMD nucleation density refers to the saturated nucleus  density, that is, the number of TMD nuclei deposited per unit of starting surface area during the steady-state growth regime after the saturation of the nucleation process, but before the coalescence \cite{routledge1970nucleation}.
This parameter can be obtained by the usual microscopic examination of the surface and is an approximation of the average crystal grain size of a fully coalesced TMD monolayer.

The degree of alignment of individual TMD nuclei is another parameter that affects intergranular defect density. Defect density decreases when individual nuclei start to align, for example,  when deposited epitaxially on a crystalline substrate \cite{peters2020directing}.
In this paper, however, we focus on the physicochemical mechanisms that determine the average crystal grain size, leaving out of scope the influence of the substrate surface topography on the degree of crystal grain alignment.

\subsection{Large spread in TMD nucleation densities and crystal grain sizes for state-of-the-art CVD chemistries: literature experimental data}

\begin{figure}
\includegraphics[width=1\columnwidth]{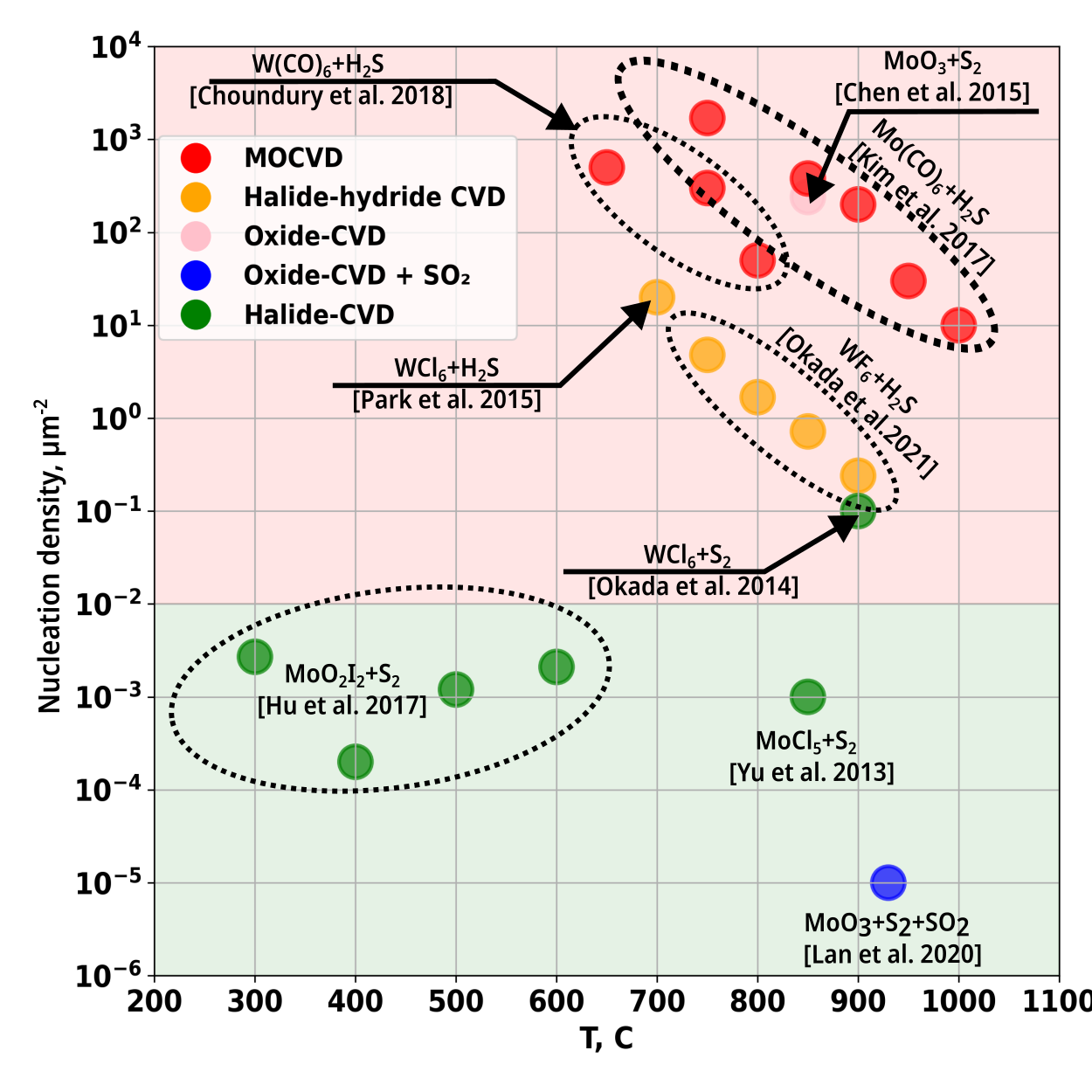}\caption{\label{fig:d_vs_t} 
Large spread in saturated TMD nucleus density for different commonly used CVD precursor chemistries:
MOCVD \cite{Kim2017,choudhury2018},
CVD from metal halide and sulfur hydride \cite{Park2015,Okada2021-WF6-H2S}
and metal oxide CVD \cite{Chen2015oxyden} exhibit high nucleation density
($10^{-1}\ldots10^{4}\,\mu{\rm m^{-2}}$), while significantly lower
densities are observed for deposition from metal halide \cite{Yu2013,Okada2014}
or oxihalide \cite{Hu2017} in combiantion with elemental sulfur, and for $\rm SO_2$-assisited metal-oxide CVD \cite{Lan2020}.}
\end{figure}

The reported nucleation densities of $\rm WS_2$ and $\rm MoS_2$  \cite{Kim2017,choudhury2018, Park2015, Chen2015oxyden, Yu2013, Okada2014, Hu2017, Lan2020, Okada2021-WF6-H2S}  vary widely by at about eight orders of magnitude,  from $10^{-5}\,\mu{\rm m^{-2}}$ to $10^{3}\,\mu{\rm m^{-2}}$, between the different CVD precursor chemistries, as shown in Figure~\ref{fig:d_vs_t}. 
Although, for a given deposition process the TMD nucleation density generally decreases with increasing deposition temperature, the precursor type may have a more significant effect. 

Indeed, certain CVD methods described in literature yield nanocrystalline monolayers of $\rm MoS_2$ and $\rm WS_2$ with nucleation densities ranging  from $10^{3}\,\mu{\rm m^{-2}}$ to $1\,\mu{\rm m^{-2}}$ regardless of the starting surface (e.g., sapphire crystal or amorphous silica dielectric).
These include 
CVD from metal oxide and elemental sulfur precursors \cite{Chen2015oxyden}, 
CVD from metal-organic and sulfur hydride precursors (MOCVD) \cite{Chung1998,Carmalt2003,mun2016low,Cadot2017,Kim2017,kim2017wafer,chiappe2018layer,choudhury2018,cwik2018direct,grundmann2019h2s,shinde2019rapid,Cohen2020,seol2020high,schaefer2021carbon}, and CVD from metal halide and sulfur hydride precursors including metal chloride \cite{Arctowski1895,Imanishi1992,Keune2000,Margolin2008WClx+H2S,Huang2014,Park2015,Campbell2022WCl6-H2S},
oxychloride \cite{Kim2022MeOCl4},
and fluoride compounds \cite{lee1994preparation,Groven2019,Okada2021-WF6-H2S}. 

In contrast, iodine-assisted chemical vapor transport (CVT) exhibits TMD nucleation density below $10^{-2}\,\mu{\rm m^{-2}}$ even at deposition temperatures as low as 300\textcelsius{} \cite{Hu2017}.
Water-assisted CVT also reduces the nucleation density below $<10^{-2}\,\mu{\rm m^{-2}}$, but only at deposition temperatures above 800\textcelsius{} \cite{Sahoo2018,Zhao2019}.

In addition, multiple works report the reduction of nucleation density by introducing co-reactants \cite{zhang2022additive-assisted}.
For example, the introduction of alkali metal halide salts \cite{Lan2020,Kim2017,Kang2015mo-decomposition} or molecular oxygen \cite{Chen2015oxyden}, into the metal-oxide CVD environment allows to reduce the nucleation density to $10^{-6}\,\mu{\rm m^{-2}}$ and yield millimeter-scale crystals.
Alternatively, co-injection of water vapor in excess of the metal precursor reduces the TMD nucleation density in metal-oxide CVD \cite{Kastl2017} and MOCVD \cite{Choi2017,Cohen2020}. 

Generalized guiding principles that could explain this wide range of TMD nucleation densities and resulting crystal grain size are still lacking in the literature.
Theoretical and experimental studies of the early stages of thin film formation indicate that the saturated nucleation density is determined by multiple process parameters, including supersaturation, rates of adsorption, desorption, and surface chemical reactions, as well as depletion of the precursor concentration around growing nuclei associated with limited surface or gas phase diffusion transport \cite{pound1963condensation,routledge1970nucleation}. 
So far however, it remains unclear which of the aforementioned processes is responsible for the observed significant difference in saturated nucleation density for different TMD deposition methods.

\subsection{Growth mechanisms that influence nucleation density and crystal grain size
in 2D material deposition}

\begin{figure}
  \includegraphics[width=1\columnwidth]{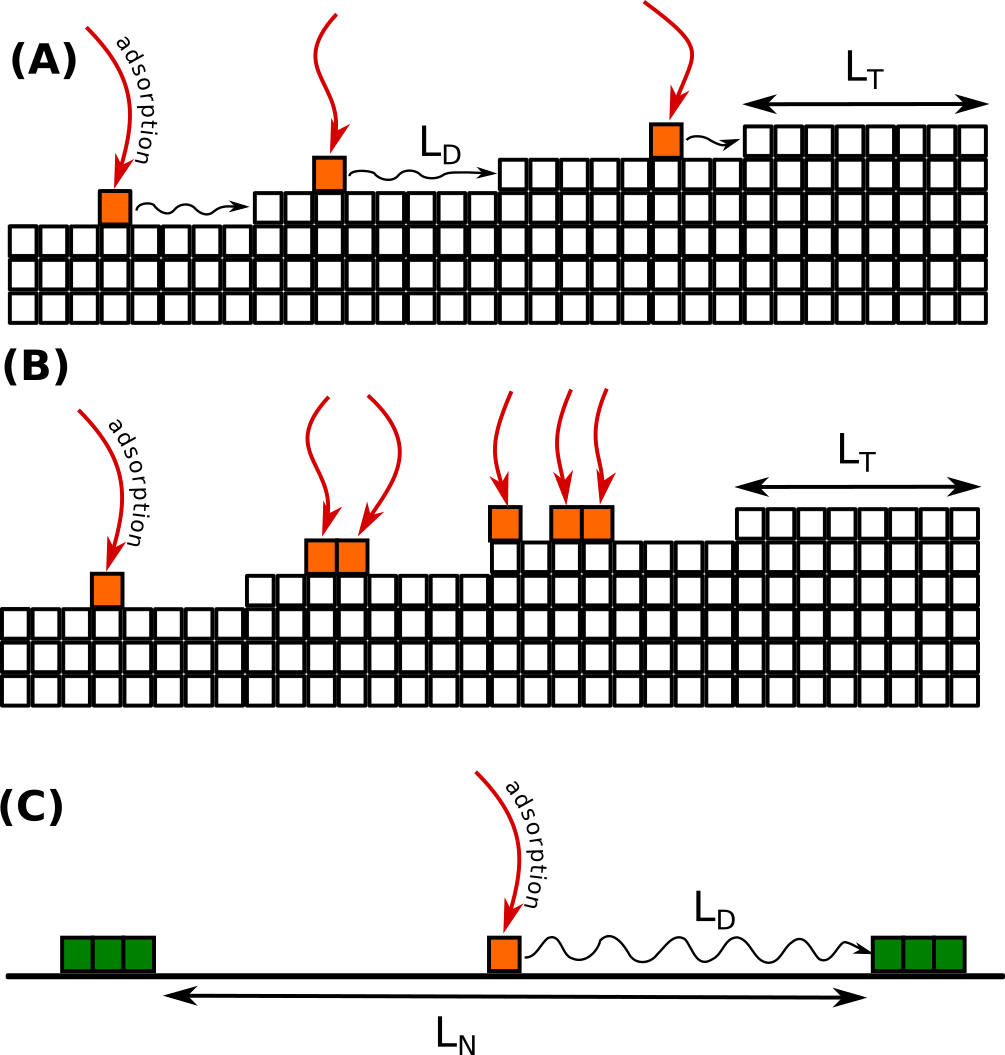}\caption{\label{fig:Epitaxial-growth}
    The influence of the lateral precursor transport and the substrate orienting effect on the crystallinity during epitaxy of 3D covalent-bonded materials and 2D van-der-Waals bonded materials.
    For conventional 3D epitaxy  on a vicinal substrate, growth proceeds in (a) step-flow mode when the precursor specie lateral diffusion length  $L_{D}$ is  larger than the terrace width  $L_{T}$ , or in (b) 3D island growth mode   when $L_{D}\ll L_{T}$ and secondary nucleation occurs on the surface of the terraces. In both cases the epilayer preserves the crystallinity of the substrate.
In contrast, to achieve a lateral epitaxial growth of 2D van-der-Walls-bonded material from 2D nuclei on a weakly bonded substrate (c), the lateral diffusion length should be larger than the average distance between the nuclei $L_{D}\gtrsim L_{N}$ to avoid secondary nucleation and prevent intergranular defect formation.}
\end{figure}

The experimentally observed spread in crystal grain size and nucleation density of $\rm MoS_2$ and $\rm WS_2$ monolayers reflects significant differences in the lateral diffusion length  of the metal precursor species, as we will show in this manuscript.
Indeed, from classical nucleation theory, the magnitude of the saturated nucleation density is proportional to the mean diffusion length of the precursor species $L_D$ \cite{bloem1977nucleation}.

As noted above, the crystal grain size in a coalesced TMD monolayer depends not only on the density of the nuclei, but also on the degree of their orientation.

If the nuclei are perfectly mutually aligned due to epitaxial interaction with the substrate, such nuclei can coalesce without intergranular boundary formation. In this case, the size of the  monocrystalline grains in the coalesced layer is much larger than the distance between the nuclei. A classic example is the homoepitaxial growth of silicon on a dislocation-free substrate, where the grain size can be as large as the substrate diameter.

In the opposite case, the interaction with the substrate is so weak that all the nuclei are randomly oriented and an intergranular boundary is formed between coalescing nuclei. Here the monocrystalline grains in the coalesced film are of the same size as the distance between the nuclei.

The Figure \ref{fig:Epitaxial-growth} illustrates how the interaction with the substrate and the magnitude of the lateral diffusion length affect the crystallinity for the case of 3D epitaxy of a covalently bonded material, such as silicon, and for the case of 2D growth of van-der-Waals bonded TMD material.

For the conventional epitaxial deposition of a single-crystal 3D material such as Si, the requirements set on the precursor diffusion length are not stringent, as such epitaxial processes rely on crystal lattice matching to the starting surface of a native or non-native single-crystal template by covalent bonding. 
For the classic example of Si homoepitaxial deposition, the crystallinity and the step-and-terrace structure of the substrate are preserved if the surface diffusion length $L_D$ is comparable to the terrace width $L_T$  $$L_D \gtrsim L_T $$ as shown in Figure \ref{fig:Epitaxial-growth}a.
Even more, monocrystalline growth of a covalently bonded crystal is possible even when the diffusion length is shorter than the terrace width $$L_D \ll L_T $$ as shown in Figure \ref{fig:Epitaxial-growth}b.
The step-and-terrace surface structure is lost due to the secondary nucleation on the surface of the terraces, but the single-crystallinity of the growing layer is preserved due to the orienting effect of the template.

 The impact of lateral diffusion is significantly stronger in the epitaxy of TMD materials, which is performed on a weakly bound substrate starting from 2D seed crystals (Figure~\ref{fig:Epitaxial-growth}c).
 Due to the van der Waals nature of TMDs, the starting surface does not \textit{a priori} exert such a strong orienting effect on the adsorbed metal and chalcogen precursor species and the 2D nuclei as in the case of homoepitaxial deposition of covalently bonded materials (e.g., Si).
Furthermore, TMD crystals preferentially grow laterally along the crystal edges of a given number of TMD nuclei until the crystals coalesce during the steady-state regime.
Hence, the TMD crystal grain size is largely determined by the density of TMD nuclei. 
The diffusion length should be greater than the inter-nucleus distance $L_N$
$$L_D \gtrsim L_N $$
in order to prevent secondary nucleation.

Thus, in van-der-Waals epitaxial growth of 2D materials, the lateral diffusion of precursor species plays a dominant role in defining the crystallinity of the deposited material.

This scientific problem of depositing large 2D crystals can be treated similar to classical area-selective deposition and epitaxial lateral overgrowth studies \cite{yamaguchi1993lateral,coltrin2003mass,shioda2009selective}.
Both cases require a complete suppression of deposition over the ``non-growth'' region of the substrate surface, \textit{i.e.}:
\begin{enumerate}
\item  masked region in the case of area-selective deposition;
  \item substrate surface between the nuclei  in the case of 2D epitaxy.
  \end{enumerate}
  Also, both cases require efficient lateral transport of precursor species to the ``growth'' regions of the substrate, \textit{i.e.}:
\begin{enumerate}
\item mask opening in the case of area-selective deposition;
  \item TMD nuclei in the case of 2D epitaxy. 
\end{enumerate}
The challenging part is that the distance between these ``growth'' regions can be as large as hundreds of microns  \cite{Gupta1999ELOG}. 

These selective deposition studies have identified two basic mechanisms involved in the lateral transport of precursor species across the starting surface: surface diffusion and gas phase diffusion \cite{yamaguchi1993lateral,OLSSON200624}. 
Typically, surface diffusion proceeds over submicron distances \cite{yamaguchi1993lateral}, and slows down exponentially with decreasing deposition temperature \cite{Naumovets1985}.
If the diffusive transport of precursor species is predominantly by surface diffusion, the deposition of a highly crystalline material at low temperatures becomes highly challenging. 

In contrast, lateral gas-phase diffusion of precursor species allows lateral mass transport over distances beyond tens of microns and is only weakly dependent on the deposition temperature.
Hence, lateral gas-phase diffusion is an essential prerequisite for any deposition process when large characteristic distances need to be overcome.
Efficient lateral gas-phase diffusion requires the use of highly volatile precursors as shown in  \cite{yamaguchi1993lateral,bradbury1984control,tsuchiya1999cl},
but quantitative criteria for the volatility of the precursors are not specified. 

Therefore, this paper defines quantitative requirements for the volatility of the deposition reaction components, that allow long-range lateral transport via gas-phase diffusion, and provides an exemplary case for the MoS\textsubscript{2} and WS\textsubscript{2} deposition. 

\subsection{Paper outline}

This paper describes the general guidelines for designing a TMD CVD process:

\begin{enumerate}
\item The nucleation density decreases to $\rm 10^{-6} \mu m^{-2}$ and below when lateral gas-phase diffusion of precursor species is possible in addition to surface diffusion.
\item  Lateral gas-phase diffusion of the metal precursor is possible when the metal precursor desorption rate  exceeds the net deposition rate, \textit{i.e.}, when the metal precursor  adsorption becomes reversible. 
\item The magnitude of the metal precursor desorption flux is estimated by thermodynamic calculation of the equilibrium partial pressure of the precursor.
  
\end{enumerate}

The paper consists of  three sections.
In section \ref{sec:The-concept}, we demonstrate how to estimate the extent of lateral gas-phase precursor diffusion for any given set of CVD precursors by calculating the precursor desorption rate using chemical equilibrium analysis.
Then, in section \ref{sec:Comparison-of-known}, we apply the proposed analytical model to the most commonly reported CVD precursor chemistries for both MoS$_2$ and WS$_2$, and predict the deposition temperature window in which lateral gas-phase diffusion occurs.
Finally, in section \ref{sec:Comparison-experiment}, we compare the predictions of the proposed model with the experimental data from the literature. The comparison confirms that reversible sorption of the metal precursor is a sufficient condition to ensure lateral gas-phase diffusion transport  and to deposit  millimeter-sized single crystals of TMD.

\section{The concept: promote reversible metal precursor sorption to enable lateral gas-phase diffusion and increase TMD crystal size \label{sec:The-concept}}

\subsection{Lateral gas phase diffusion is possible only if desorption rate is  higher than net deposition rate}

The TMD deposition process involves the transport of both metal and chalcogen precursors. However, under typical TMD deposition conditions, the chalcogen precursor concentration is several orders of magnitude higher than the metal precursor concentration. As a result, the TMD nucleation and growth processes are limited by the metal precursor. Therefore, in the remainder of this paper, we will focus only on the transport of the metal precursor.

\begin{figure}
  \includegraphics[width=1\columnwidth]{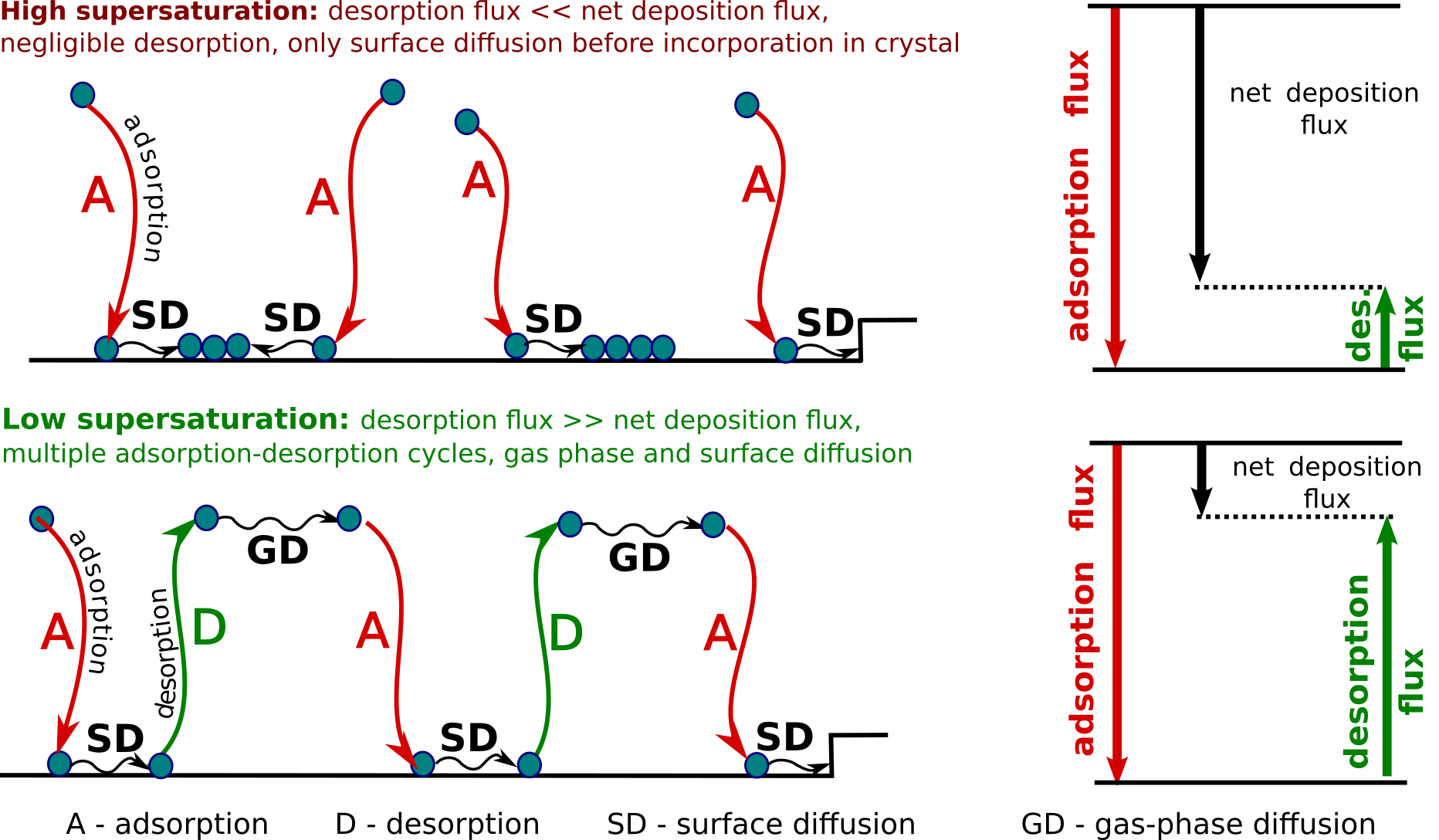}\caption{\label{fig:ad-des}
For high supersaturation deposition conditions (a), the desorption flux of the metal precursor is negligible  ($J_{des}\ll J_{net}$).
The adsorption of the metal precursor species is irreversible, and the only
lateral transport mechanism is surface diffusion.
For  low supersaturation deposition conditions (b), the metal precursor desorption flux exceeds the net deposition flux ($J_{des}\gg J_{net}$).
Metal precursor species undergo multiple successive cycles of adsorption
and desorption, making gas-phase diffusion the dominant mechanism
of lateral transport.}
\end{figure}

Efficient lateral gas-phase transport relates to the reversibility of metal precursor sorption on the starting surface, as shown schematically in Figure \ref{fig:ad-des}.

In the limiting case of negligibly small desorption (Figure \ref{fig:ad-des}a), any precursor species colliding with the surface will stick to the surface and never leave it. In this case, the only mechanism for lateral mobility of adsorbed precursor species is surface diffusion. 

In the opposite case (Figure \ref{fig:ad-des}b), the adsorption and desorption fluxes are nearly equal, so a precursor atom undergoes multiple cycles of adsorption, surface diffusion, desorption, and gas-phase diffusion before being incorporated into the growing crystal.
Thus, gas-phase diffusion contributes  significantly to the lateral transport when the desorption flux $J_{des}$ is much larger than the net deposition
flux $J_{net}$:

\begin{equation}
J_{des}\gg J_{net}\label{eq:criterion}
\end{equation}

\subsection{What mechanisms control the desorption rate?}

Let's consider how to design a deposition process that satisfies this condition.

This section examines two limiting cases:
\begin{enumerate}
\item deposition rate is limited by the kinetics of the surface reactions;
\item deposition process has infinitely fast surface kinetics.
\end{enumerate}
In the limiting case of surface-kinetics-limited deposition, the partial pressure of the metal precursor at the substrate surface  $P_{Me}^{surf}$ is equal to the initial partial pressure of the metal precursor
$P_{Me}^{i}$:

\begin{equation}
  \label{eq:p_in_kin_limit}
P_{Me}^{surf}\Bigg|_{{\rm slow\,surface\,reaction}}=P_{Me}^{i}
\end{equation}

In this case, the net deposition flux is, by definition, much smaller than the adsorption and desorption fluxes, which can be calculated using the Hertz-Knudsen equation \cite{pound1963condensation}:

\begin{equation}
J_{net}\ll J_{des}^{kin}\approx J_{ads}^{kin}=\frac{P_{Me}^{i}}{\sqrt{2\pi m_{Me}k_{B}T}}\label{eq:des_flux_kin}
\end{equation}
where $m_{Me}$ is mass of metal precursor specie, $T$ is temperature, and $k_{B}$ is Boltzmann constant.

In the opposite case of infinitely fast surface reaction kinetics, the deposition reaction at the substrate surface is close to equilibrium, and the metal precursor partial pressure is close to the equilibrium value
$P_{Me}^{eq}$:

\begin{equation}
P_{Me}^{surf}\Bigg|_{{\rm fast\,surface\,reaction}}=P_{Me}^{eq}
\end{equation}

As a result, the gas mixture near the surface is significantly depleted of metal precursor and the net condensation flux is determined by the diffusion through the boundary layer \cite{Goodman1974}:

\begin{equation}
J_{net}\approx J_{net}^{diff}=D_{Me}\frac{P_{Me}^{i}-P_{Me}^{eq}}{\delta}\label{eq:diff-lim}
\end{equation}
where $D_{Me}$ is the gas phase diffusion coefficient of the metal precursor species, and $\delta$ is the boundary layer thickness. The metal precursor desorption flux is proportional to $P_{Me}^{eq}$:

\begin{equation}
J_{des}^{eq}=\frac{P_{Me}^{eq}}{\sqrt{2\pi mk_{B}T}}\label{eq:des_flux_eq}
\end{equation}

In all intermediate cases, the actual metal precursor desorption flux
is in the range from $J_{des}^{eq}$ to $J_{des}^{kin}$:

\begin{equation}
\underset{{\rm Fast}\xleftarrow[{\rm }]{{\rm surface\,kinetics}}{\rm Slow}}{J_{des}^{eq}<J_{des}<J_{des}^{kin}}\label{eq:J-interval}
\end{equation}

Now, based on the Equation~\ref{eq:J-interval}, there are two opposing approaches to deliberately increase the metal precursor desorption flux:

The first approach is to keep the surface reaction away from the reaction equilibrium. 
The further away from equilibrium the surface reaction is, the higher the $P_{Me}^{surf}$ and thus the metal precursor desorption flux becomes. Ultimately, the metal precursor desorption flux can be increased up to $J_{des}^{kin}$ according to Equation~\ref{eq:des_flux_kin}.
This can be achieved, for example, by lowering the deposition temperature
in order to slow down the rate of the surface reaction. 

Another way to slow down the deposition kinetics is to reduce the nucleation rate of TMD crystals on the substrate by reducing supersaturation.
For example, if the precursor supersaturation corresponds to the Ostwald-Meyers metastable region, the nucleation will be completely inhibited. In the absence of pre-existing ``seed'' crystals the net deposition rate is zero. 

In the second approach, the surface reaction proceeds close to equilibrium so that the metal precursor desorption flux $J_{des}^{eq}$ is proportional the equilibrium metal precursor pressure $P_{Me}^{eq}$ (Equation~\ref{eq:des_flux_eq}).
The magnitude of the equilibrium pressure $P_{Me}^{eq}$ for a given precursor can be controlled not only by the deposition temperature, but also by the composition of the gas mixture.
This provides more degrees of freedom in the design of the deposition process.
Following Eqs.~\ref{eq:criterion} and \ref{eq:des_flux_eq} we can estimate the range of metal precursor equilibrium pressures at which the gas phase lateral diffusion transport is activated for a given deposition rate:

\begin{equation}
P_{Me}^{eq}\gg J_{net}\sqrt{2\pi mk_{B}T}
\end{equation}

In this paper, we will focus on the second approach, as low supersaturation usually favors the growth of pristine crystals.

\section{Type of CVD precursor couple determines equilibrium metal precursor
desorption rate\label{sec:Comparison-of-known}}

\subsection{Calculating equilibrium metal precursor desorption flux}

\begin{figure*}
\includegraphics[width=1.0\textwidth]{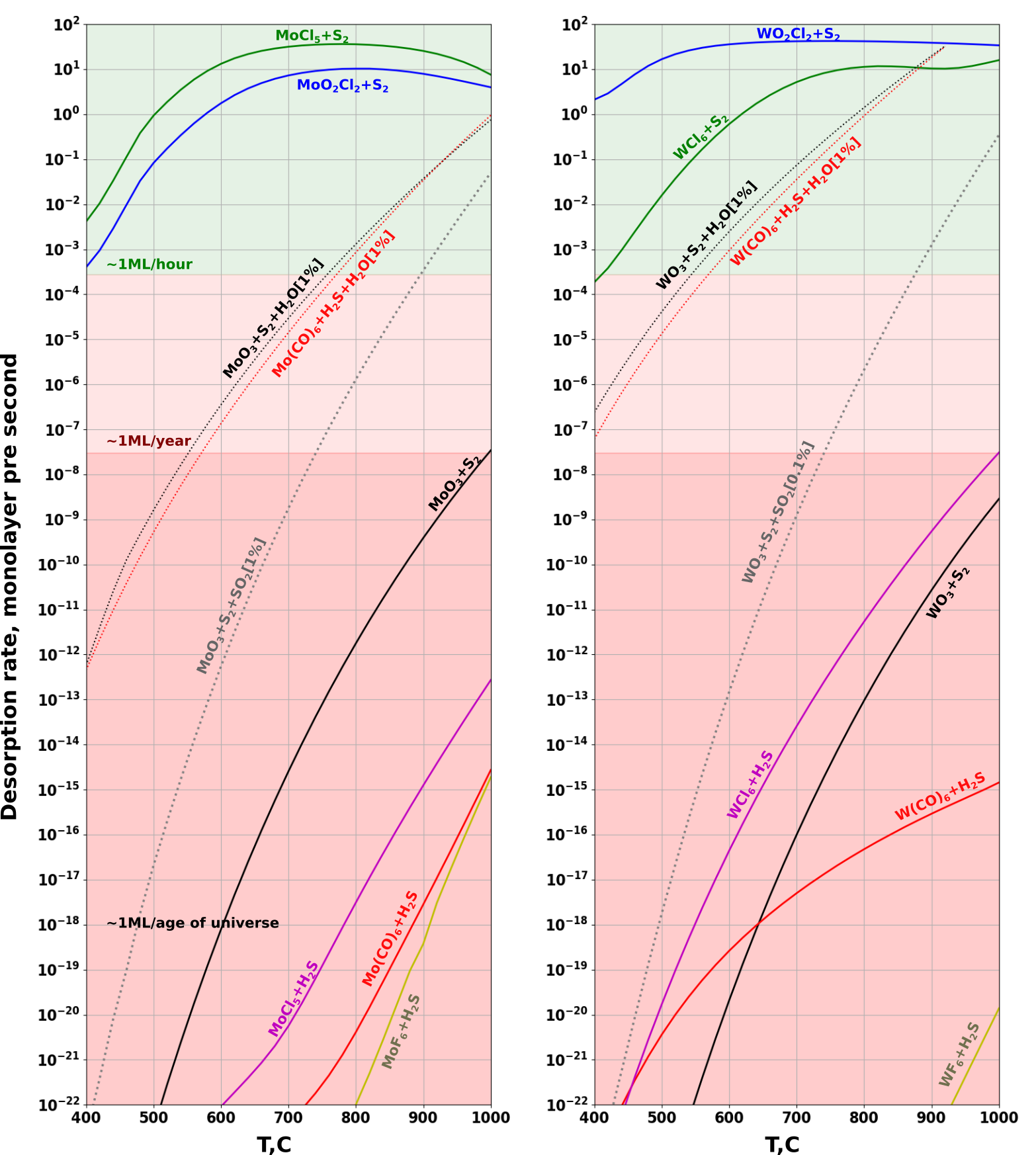}
\caption{\label{fig:Equilibrium-desorption-rate} 
  Equilibrium desorption rate of metal precursor species over condensed MoS\protect\textsubscript{2} (left) and WS\protect\textsubscript{2} (right) for different combinations of CVD precursors and co-reactants, estimated from the equilibrium calculation of the gas-phase composition using the Hertz-Knudsen equation.
  Several methods, including MOCVD, halide-hydride CVD, and oxide-CVD demonstrate negligibly low equilibrium desorption rates in all investigated temperature range.
  In contrast, deposition from metal halide or oxihalide and elemental sulfur exhibit desorption rates higher than 1 monolayer per second for temperatures above 600\textcelsius.
The addition of water or sulfur oxide to the gas mixture increases the desorption rate by several orders of magnitude.}
\end{figure*}

Next, the values of the  equilibrium metal precursor desorption flux are calculated for several common deposition methods (Figure \ref{fig:Equilibrium-desorption-rate}).
The details of the calculation procedure are given in section \ref{app:thermo}.
All calculations assume a metal precursor molar concentration of $10^{-6}$, a sulfur precursor molar  concentration of $10^{-3}$, and a total pressure of 1~atm, that is, the conditions that are representative of a typical TMD CVD process.

\subsection{TMD deposition methods demonstrating high equilibrium metal precursor desorption rate\label{subsec:Methods-with-high}}

The deposition of MoS$_2$ and WS$_2$ from a metal halide and elemental sulfur proceeds through a group of deposition reactions 
\begin{equation}
{\rm MoCl_{4}^{g}+S_{2}^{g}\leftrightarrow MoS_{2}^{s}+SCl_{2}^{g}/S_{2}Cl^{g}/Cl^{g}}\label{eq:MoCl-S}
\end{equation}
and 
\begin{equation}
{\rm WCl_{6}^{g}+S_{2}^{g}\leftrightarrow WS_{2}+SCl_{2}^g/S_{2}Cl^g/Cl^g}\label{eq:WCl-S}
\end{equation}
respectively, where the aggregate state of the substance is indicated by a superscript: $ s$ for solid and $g$ for gas. 
This system exhibits an equilibrium metal precursor desorption rate exceeding 1~monolayer (ML) per second at deposition temperatures higher than 500\textcelsius{} for MoS\textsubscript{2} deposition and 600\textcelsius{} for WS\textsubscript{2} deposition.

Similarly, when employing a metal oxyhalide precursor instead of a metal halide precursor in combination with elemental sulfur, the overall deposition proceeds according to reactions 
\begin{equation}
{\rm MoO_{2}Cl_{2}^{g}+S_{2}^{g}\leftrightarrow MoS_{2}^{s}+S_{2}Cl^{g}/Cl^{g}+SO_{2}^{g}}\label{eq:MoOCl-S}
\end{equation}
 and 
\begin{equation}
{\rm WO_{2}Cl_{2}^{g}+S_{2}^{g}\leftrightarrow WS_{2}^{s}+S_{2}Cl^{g}/Cl^{g}+SO_{2}^{g}}\label{eq:WOCl-S}
\end{equation}
and the equilibrium metal precursor desorption rate exceeds 1~ML per second at deposition temperatures above 600\textcelsius{} for MoS$_2$ and over the entire deposition temperature window studied for WS$_2$.

In should be noted, that the dominant deposition product for reactions (\ref{eq:MoCl-S}\ldots\ref{eq:WOCl-S}) depends on temperature: sulfur chlorides prevail at deposition temperatures below 800\textcelsius{}, while chlorine dominates at higher temperatures.

\subsection{TMD deposition methods demonstrating negligibly low equilibrium metal
precursor desorption rate}

In contrast, CVD of MoS$_2$ and WS$_2$ from a metal-organic or metal oxide precursor, regardless of the  sulfur precursor type, as well as CVD from a metal halide and sulfur hydride  exhibit negligibly low equilibrium metal precursor desorption rates, i.e., less than 1ML/year at deposition temperatures up to 1000\textcelsius{}.
However, all of these deposition chemistries are highly sensitive to the injection of additives such as molecular oxygen, water, and alkali metal halides, as discussed in the section \ref{subsec:Additives-improving-volatility}.

\subsection{Additives, that increase equilibrium desorption rate \label{subsec:Additives-improving-volatility}}

The introduction of certain additives, such as  alkali metal halides, molecular oxygen, or water vapor, significantly reduces the TMD nucleation rate and thus increases the crystal grain size.
Here we categorize two types of such additives:
\begin{enumerate}
\item  gaseous by-products of the deposition reaction
  \item co-reactants that react with the metal precursor.
\end{enumerate} 

\paragraph{Gaseous by-products of the TMD deposition reaction}

This type of additive does not alter the deposition reaction, but rather shifts the equilibrium of the deposition reaction toward the side of the reactants (i.e., away from the reaction products).
An example of such an additive is sulfur dioxide ($\rm SO_2$), which is produced during metal oxide CVD.

\paragraph{Co-reactants}

These substances react with the metal precursor to form a more volatile metal compound.
In this case, TMD deposition occurs via a different deposition reaction, which has a higher equilibrium metal precursor partial pressure and thus a higher equilibrium desorption rate.
Typical examples of such co-reactants are halogen-based compounds (e.g., NaCl) and water vapor.
These co-reactants are co-injected with the carrier gas during metal-oxide or metal-organic CVD.

Note that a significant portion of the literature is devoted to vapor-liquid-solid (VLS) growth on substrates containing a predetermined amount of condensed alkali metal halide compounds \cite{Yu2020vls}. However, this interesting approach is beyond the scope of the present research, since the transfer of the precursor to the growth front occurs in the condensed phase, and not by surface or gas-phase diffusion.

\subsubsection{Sulfur dioxide as a by-product during metal oxide CVD\label{subsec:Sulfur-dioxide-as}}

Sulfur dioxide ($\rm SO_2$) forms as a by-product during the CVD of MoS$_2$ and WS$_2$  from a metal oxide and elemental sulfur precursor, according to the overall deposition reactions (Eqs.~\ref{eq:MoO-sulfurization} and
\ref{eq:WO-sulfurization}). 

\begin{equation}
{\rm MoO_{3}^{g}+\frac{7}{4}S_{2}^{g}\rightarrow MoS_{2}^{s}+\frac{3}{2}SO_{2}^{g}}\label{eq:MoO-sulfurization}
\end{equation}

\begin{equation}
{\rm WO_{3}^{g}+\frac{7}{4}S_{2}^{g}\rightarrow WS_{2}^{s}+\frac{3}{2}SO_{2}^{g}}\label{eq:WO-sulfurization}
\end{equation}

Thus, intentional co-injection of  $\rm SO_2$ shifts the reaction equilibrium to the left-hand side (i.e., the side of the reactants), thereby increasing the equilibrium partial pressure of the metal oxide precursor proportional to:

$$
P_{{\rm MoO_{3}}}^{eq}\propto P_{{\rm SO_{2}}}^{\frac{3}{2}}
$$

For example, the addition of 1\% molar percent of $\rm SO_2$ increases the equilibrium metal oxide precursor desorption rate for both Mo and W by six orders of magnitude (Figure \ref{fig:Equilibrium-desorption-rate}).

To the best of our knowledge, experiments in which a fixed and controlled amount of gaseous $\rm SO_2$ is injected intentionally and simultaneously into the CVD gas mixture have not been reported in literature.
However, $\rm SO_2$ forms as a by-product during the metal oxide CVD process, for example when sulfur reacts with metal oxide precursor or gaseous molecular oxygen via reactions \ref{eq:MoO-sulfurization} and \ref{eq:WO-sulfurization}:
\begin{itemize}
\item In typical oxide-CVD experimental setups, \cite{lee2012synthesis,Hyun2018MoO3-S2,Lan2020} sulfur vapor get in contact with a metal oxide source, producing sulfur dioxide. Evidence of this reaction is the significant conversion of metal oxide to lower oxides during the deposition \cite{Hyun2018MoO3-S2}. 
\item Another way to produce SO\textsubscript{2} is to add molecular oxygen to the carrier gas \cite{Chen2015oxyden}, where the elemental sulfur reacts with the oxygen to produce sulfur dioxide:
\end{itemize}
$$
{\rm S_{2}+2O_{2}\rightarrow2SO_{2}}
$$

\subsubsection{Alkali metal halide as a co-reactant during metal oxide CVD \label{subsec:Halide-assisted-Oxide-CVD}}

When alkali metal halides are added to the crucible containing the metal oxide precursor, both compounds react producing a more volatile metal oxychloride \cite{Johnson1982moo3_NaCl} according to reactions:
\begin{equation}
  {\rm NaCl^{s}+3MoO_{3}^{s}\rightarrow MoO_{2}Cl_{2}^{g}+Na_{2}Mo_{2}O_{7}}^{{\rm s,l}}\label{eq:MoO3-NaCl}
  \end{equation}
  
  \begin{equation}
  {\rm NaCl^{s}+3WO_{3}^{s}\rightarrow WO_{2}Cl_{2}^{g}+Na_{2}W_{2}O_{7}}^{{\rm s,l}}\label{eq:WO3-NaCl}
  \end{equation}
As a result, the TMD deposition proceeds similarly to metal oxyhalide
CVD according to the deposition reactions \ref{eq:MoOCl-S} or \ref{eq:WOCl-S}
demonstrating high $J_{des}^{eq}$, as discussed in section \ref{subsec:Methods-with-high}.

\subsubsection{Water-assisted metal oxide and metal-organic CVD}

Co-injection of  water vapor during either metal oxide or metal-organic CVD affects the TMD deposition process in two different ways. 

First, water reacts with the initial metal precursor to form a more volatile metal compound, such as a metal hydroxide.
Such volatile hydroxides of molybdenum (MoH\textsubscript{2}O\textsubscript{4}) and tungsten (WH\textsubscript{2}O\textsubscript{4}) form in presence of water vapor by reaction with metal \cite{Belton1964,Belton1965,Schaefer1973},metal oxide \cite{Zheng2017MoO3-S2-separate,Chen2018hydrogen-assisted-MoO3-S2}, metal chalcogenide \cite{Sahoo2018,Zhao2019}, or metal-organic precursor \cite{Choi2017,Cohen2020}. 

Second, water alters the reaction equilibrium of the TMD deposition reaction as it is one of the gaseous reaction products of the reaction between MoH\textsubscript{2}O\textsubscript{4} and H\textsubscript{2}S precursors according to \ref{eq:MoH2O2}.

\begin{equation}
\rm MoH_{2}O_{4}^{g}+\frac{7}{3}H_{2}S\leftrightarrow MoS_{2}^{s}+\frac{10}{3}H_{2}O^{g}+\frac{1}{3}SO_{2}^{g}\label{eq:MoH2O2}
\end{equation}

Adding 1\% volume percent of water vapor during metal oxide or metal-organic CVD increases $J_{des}^{eq}$ to 1ML/s at deposition temperatures above 1000\textcelsius{} for MoS$_2$ and above 800\textcelsius{} for WS$_2$ deposition, respectively (Figure \ref{fig:Equilibrium-desorption-rate}).

\section{Benchmarking the model with experimental data\label{sec:Comparison-experiment}}
\subsection{Comparison to published deposition methods}
\begin{figure}[t]
\includegraphics[width=1\columnwidth]{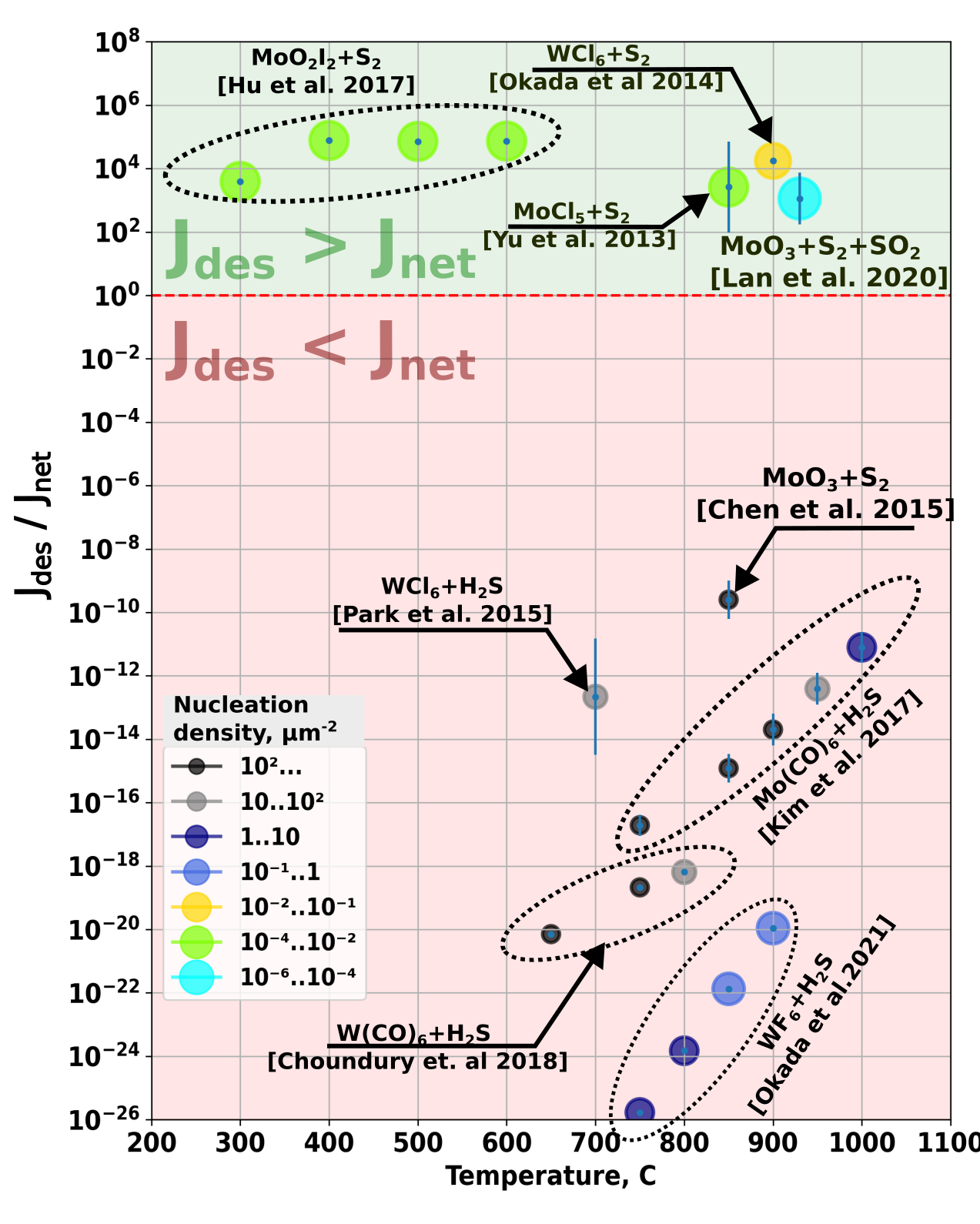}\caption{\label{fig:COMPARIZON}Dependence of nucleation density on temperature and ratio $\nicefrac{J_{des}}{J_{net}}$ for experimental data reported in \cite{Kim2017,choudhury2018, Park2015, Chen2015oxyden, Yu2013, Okada2014, Hu2017, Lan2020, Okada2021-WF6-H2S}}
\end{figure}

This last section compares the above calculated values of the equilibrium metal precursor desorption flux with the available experimental data on the TMD nucleation density and crystal grain size for the most reported TMD deposition methods \cite{Yu2013,Okada2014,Chen2015oxyden,Park2015,Hu2017,Kim2017,choudhury2018,Lan2020,Okada2021-WF6-H2S}.

For each literature source, the value of $J_{des}^{eq}$ was estimated, using the experimental details provided by the respective authors: this includes deposition temperature, total pressure, initial composition of the precursor mixture supplied to the reactor, and resulting TMD deposition rate.
This comparison with the literature reveals how the TMD nucleation density, and hence the crystal grain size, depends on the deposition temperature and on the estimated ratio of metal precursor desorption flux to net deposition flux $J_{des}^{eq}/J_{net}$ (Figure \ref{fig:COMPARIZON}).

Indeed, when the condition $J_{des}^{eq}>J_{net}$ is satisfied, the TMD nucleation density ranges from $10^{-1}\mu{\rm m}^{-2}$\cite{Okada2014} down to $10^{-6}\mu{\rm m}^{-2}$ \cite{Lan2020} and the crystal grain size increases above $100\,\mu{\rm m}$ \cite{Lan2020} for metal halide\cite{Yu2013,Okada2014}, metal oxyhalide \cite{Hu2017} and metal oxide CVD \cite{Lan2020}.
In the latter case, however, SO\textsubscript{2} is formed as a reaction by-product from the reaction of the metal oxide with the sulfur precursor, as described in section \ref{subsec:Sulfur-dioxide-as}. 

In contrast, for deposition methods where the equilibrium metal precursor desorption flux is negligibly small, the TMD nucleation density remains high between $1\,\mu{\rm m}^{-2}$ and $10^{3}\mu{\rm m}^{-2}$, and the representative crystal grain size falls into the submicron range.
As detailed in section \ref{sec:Comparison-of-known}, these deposition methods include MOCVD \cite{Kim2017,choudhury2018},  CVD from metal halide and sulfur hydride precursors \cite{Park2015,Okada2021-WF6-H2S} and metal oxide CVD \cite{Chen2015oxyden}.

\begin{figure*}[ht!]
\includegraphics[width=1\textwidth]{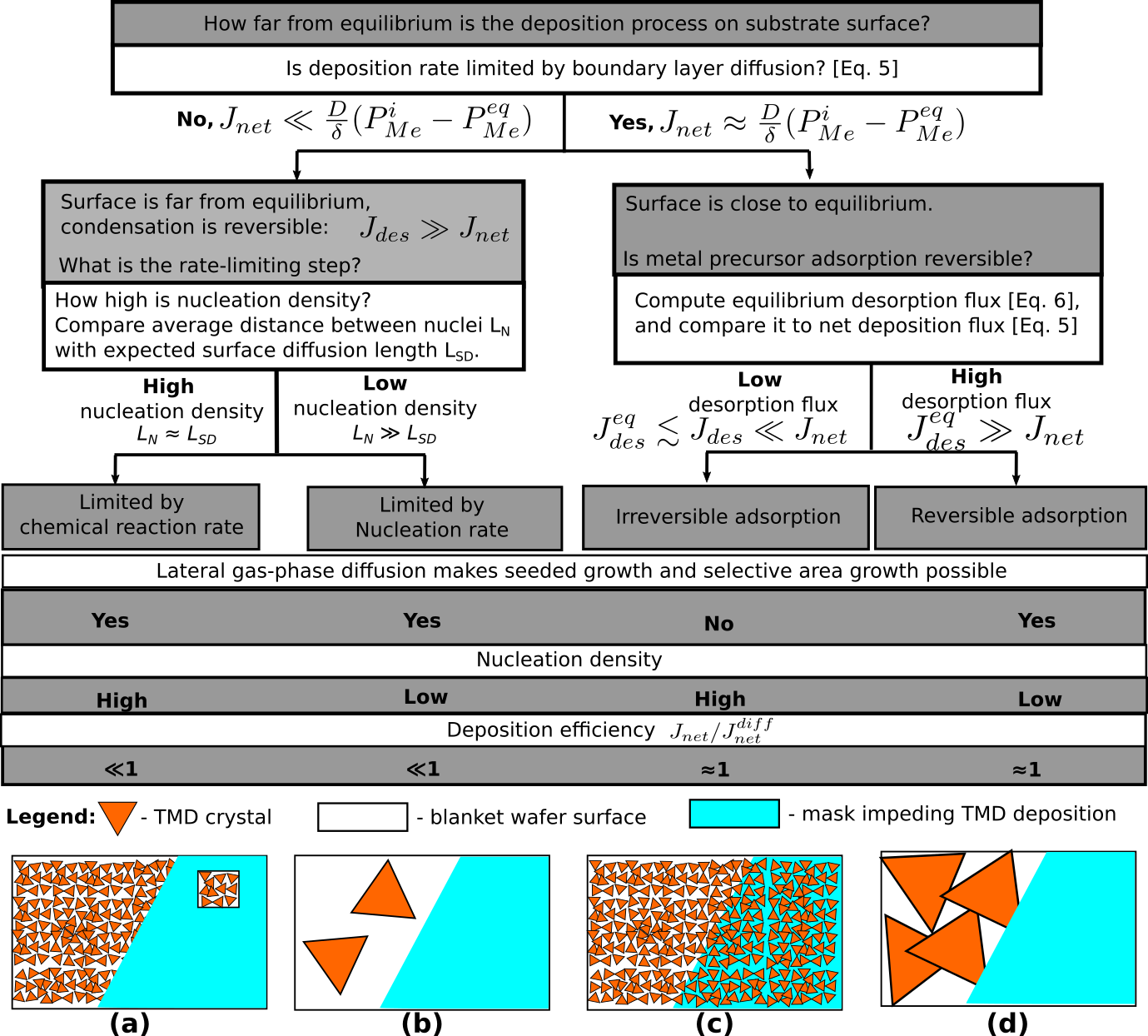}\caption{\label{fig:classify}
The calculated equilibrium desorption rate and the experimentally observed deposition rate and material crystallinity help to determine the rate-limiting step and the metal precursor desorption rate.
(a,b) The deposition rate is limited by surface reaction kinetics, with the surface far from equilibrium. The type of limiting kinetic step can be inferred from the TMD nucleation density: (a) surface reaction limited process -- high nucleation density, selective deposition possible using masks that block the surface reaction process; (b) nucleation rate limited process --  low nucleation density. (c,d) Diffusion-limited deposition rate, with the surface near equilibrium: (c) negligible equilibrium metal precursor desorption flux -- high nucleation density, non-selective deposition (d)  equilibrium desorption flux greater  than the net deposition flux -- low nucleation density, selective deposition possible.}
\end{figure*}

\subsection{Limits of applicability: how to determine
true value of metal precursor desorption rate?}

\subsubsection{Complementing model predictions  with experimental data}

The equilibrium desorption flux calculated using Equation.~\ref{eq:des_flux_eq} represents the lower limit of the actual desorption flux. The latter can differ significantly from its equilibrium value depending on the rate of surface kinetics. The influence of surface kinetics can be estimated from experimentally determined deposition rates and nucleation densities  as summarized in Figure~\ref{fig:classify}.

The type of rate-limiting process can be identified  by measuring the  TMD deposition rate $J_{net}$ and comparing it with the value of the deposition rate in the diffusion boundary layer limit $J^{diff}_{net}$  calculated by Equation~\ref{eq:diff-lim}:
\begin{enumerate}
	     \item  Limiting step is one of the surface processes (Figure~\ref{fig:classify}a,b) when $J_{net}\ll J^{diff}_{net}$.
     \item  Deposition is limited by  diffusion through the boundary layer  (Figure~\ref{fig:classify}c,d) when $J_{net}\ll J^{diff}_{net}$.
\end{enumerate}

An assumption about the nature of the limiting stage of the kinetics can be made based on the measured nucleation density of the deposit and a thermochemical estimate of the supersaturation value.

Below we consider typical cases of deposition processes limited by surface kinetics and by boundary-layer diffusion. 

  \subsubsection{Surface kinetics limited deposition}
When the deposition rate is limited by surface kinetics, the  gas phase is not depleted of metal precursor (Equation~\ref{eq:p_in_kin_limit}). As a result, the lateral gas-phase diffusion mechanism is active since the condition for reversible sorption (Equation~\ref{eq:criterion}) is satisfied.  The deposition reaction at the surface is far from equilibrium under these conditions.

The process limiting the deposition rate can be either an elementary microscopic process, such as a chemical reaction, or a nucleation phenomenon. These cases can be distinguished by the observed nucleation density:
\begin{enumerate}
  
\item Deposition limited by the surface chemical reaction rate under high supersaturation conditions will exhibit a high nucleation density,resulting in a polycrystalline structure of the deposited TMD material (Figure~\ref{fig:classify}a). At the same time, efficient long-range lateral transport via gas-phase diffusion can proceed since the adsorption is reversible.

  The features of this deposition regime can be illustrated by the area-selective deposition experiments.  
    On the one hand, area-selective deposition is achieved despite high supersaturation when the deposition reaction is suppressed on a certain area of the substrate by applying a mask or surface functionalization \cite{Bersch2017,Ahn2021}. Deposition on masks is suppressed even if the masked areas are much larger than the surface diffusion length.
     On the other hand, the crystal perfection and purity of the material deposited inside the mask openings are usually poor \cite{kim2017grains}.

\item Deposition limited by the nucleation kinetics can be experimentally identified by the dependence of the deposition rate on the areal density of seed crystals on the substrate surface (Figure \ref{fig:classify}b). In the Ostwald-Meyers region, the deposition occurs exclusively on the seed crystals present on the substrate surface. The low nucleation density determines the larger achievable size of monocrystalline grains compared to kinetically limited deposition at high supersaturation.
\end{enumerate}

It should be noted that these two cases are not mutually exclusive.
For example,  if the surface reaction rate is zero, then the nucleation rate is also zero.
Consider the opposite case: low supersaturation and therefore low nucleation density. If the total deposition rate is less than the diffusion-limited rate (Equation~\ref{eq:diff-lim}), this indicates the presence of a kinetic constraint on the lateral rate of crystallite growth. One mechanism for such a kinetic limitation may be the limited rate of chemical reaction at the edge of the crystallites.

\subsubsection{Boundary layer diffusion limited deposition}
Now consider the limiting case of infinitely fast surface reactions, so that  the actual precursor desorption flux is close to its equilibrium value (Figure~\ref{fig:classify}c-d).
For this type of processes, the model proposed in this paper is applicable.  Let us consider two limiting cases depending on the ratio of the calculated equilibrium desorption flux (Equation~ \ref{eq:des_flux_eq}) to the net deposition rate. 

If the  equilibrium desorption flux is negligible compared to the deposition rate, then the metal precursor adsorption is irreversible and the crystal grain size is determined solely by the surface diffusion length, which usually ranges from tens of nanometers to several micrometers and depends exponentially on temperature (Figure \ref{fig:classify}c). Typical examples are MOCVD \cite{Kim2017,choudhury2018} and chloride-hydride CVD \cite{Okada2021-WF6-H2S}.  

On the contraty, if the equilibrium metal precursor desorption flux exceeds the net deposition flux, the gas-phase diffusion leads to a significant increase in crystal size, as described in more detail in the  previous two sections (Figure~\ref{fig:classify}d). Examples include processes using elemental sulfur and metal halide or oxyhalide precursors \cite{Hu2017,Yu2013,Okada2014}.

\section{Conclusions}

\begin{figure*}[ht!]
  \includegraphics[width=1.0\textwidth]{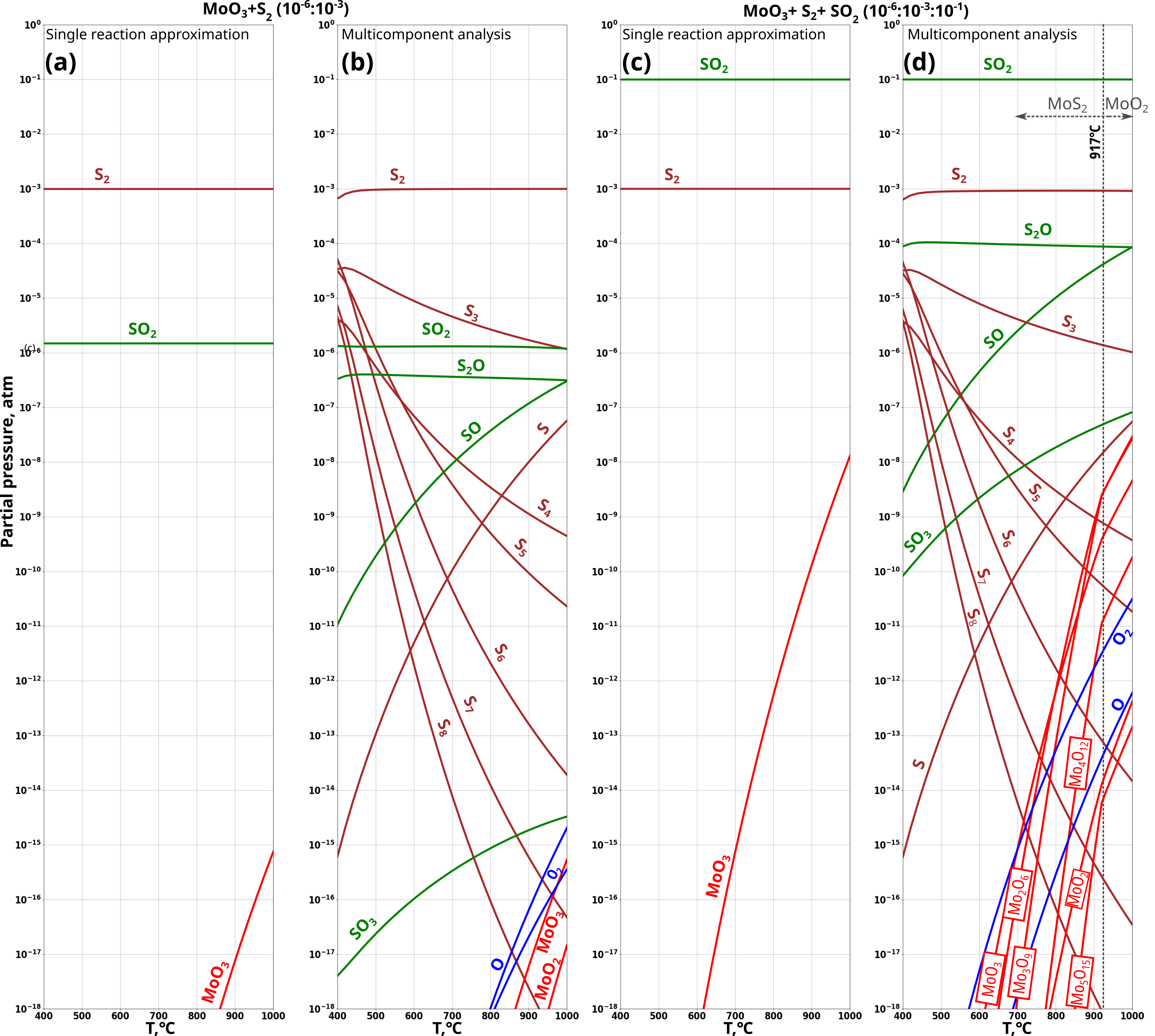}
  \caption{(a,b) Equilibrium vapor pressures of species in $\rm MoS_2$ oxide CVD calculated in single reaction approximation~(a) and by multicomponent analysis~(b).
 (c,d)    Equilibrium vapor pressures in $\rm MoS_2$ oxide CVD system  with addition of $\rm SO_2$, calculated  in single reaction approximation~(c) and by multicomponent analysis~(d). The kink in the partial pressure curves of molybdenum compounds calculated by multicomponent analysis observed at a temperature near 917\textcelsius{} is caused by a change  in the composition of the condensed phase (see gray dashed line): the equilibrium condensed  compound of molybdenum is $\rm MoS_2$ at temperatures below 917\textcelsius{} and  $\rm MoO_2$ at  temperatures above 917\textcelsius{}.
  }
  \label{fig:numeric-example}
\end{figure*}

Lateral gas-phase diffusion is an essential prerequisite to for the formation of highly crystalline TMD monolayers on amorphous substrates. It occurs over a wider deposition temperature range and larger length scales (cm-scale) than surface diffusion, yet its contribution to TMD deposition processes is often overlooked in the literature.

For reversible adsorption and lateral gas-phase diffusion to occur, the equilibrium metal precursor desorption flux should exceed the net deposition flux. Satisfying this singular condition is sufficient, as the actual metal precursor desorption flux is, by definition, always greater than the equilibrium desorption flux, even though the rate of the surface reaction kinetics is not known \textit{a priori}.

Comparison of the proposed model with experimental data from the literature confirms that the TMD nucleation density decreases and the resulting crystal size increases when the metal precursor adsorption becomes reversible, and thus allowing efficient lateral gas-phase diffusion of the metal precursor. 

Metal-halide precursors in combination with elemental sulfur provide the highest metal precursor volatility among commonly used  precursor chemistries.

Two mechanisms by which co-reagents improve TMD crystallinity are shown.
The first type of co-reagent is a gaseous by-product of the TMD deposition reaction.
The second type of co-reagent is a substance that reacts with the metal precursor to form a more volatile metal compound.

\section{Methods}
\subsection{Estimating reversability of metal precursor adsorption based on thermochemical data}

Adsorption is reversible when the metal precursor desorption flux exceeds the net deposition flux (Equation~\ref{eq:criterion}).
The desorption flux in the process depends on the rate of surface kinetics, and is not known \textit{a priori}, but it is always greater than the equilibrium desorption flux (Equation~\ref{eq:J-interval}). Therefore the equilibrium desorption flux (Equation~\ref{eq:des_flux_eq}) exceeding the net deposition flux represents a sufficient condition for reversible adsorption.

Thus, the adsorption reversibility for a given deposition process can be estimated using the following calculation procedure:
\begin{enumerate}
\item Define the process parameters required for the equilibrium calculation: the composition of the precursor mixture at the inlet of the deposition chamber, the pressure and temperature of the process.
  \item Identify potential deposition reaction products to be considered in the equilibrium calculation. 
  \item Calculate the equilibrium composition of the deposition reaction products using the numerical Gibbs free energy minimization algorithm \cite{Villars1959,Villars1960,Colonna2004,zeleznik1968calculation,zeggeren1970computation, smith-missen1983}.
  \item Calculate the net deposition flux $J^{diff}_{net}$ for the case of a boundary layer diffusion-limited deposition using Equation~\ref{eq:diff-lim}.
  \item Calculate the equilibrium desorption flux $J^{eq}_{des}$ using Equation~\ref{eq:des_flux_eq}.
  \item Check if  $J^{eq}_{des} \gg J^{diff}_{net}$.
\end{enumerate}

Several numerical examples illustrating this computational procedure are given below in the section~\ref{app:calcul}.

\subsection{Thermodynamic equilibrium calculation \label{app:thermo}}
The equilibrium vapor pressures of the reagents involved in the deposition process are evaluated by numerical minimization of the Gibbs free energy using a hierarchical algorithm \cite{Villars1959,Villars1960,Colonna2004}, while the equilibrium desorption fluxes are calculated through Equation~\ref{eq:des_flux_eq}.
Thermochemical properties of substances
from \cite{Dittmer1983,Gurvich1990,barin1995,Chase1998} are used.
The substances considered in the calculation are listed below, where superscript indicates the phase of the substance: \textit{s}~--~solid, \textit{l}~--~liquid, and \textit{g}~--~gaseous:

\paragraph*{Molybdenum}
${\rm Mo^{s}}$, ${\rm MoO_{2}^{s}}$, ${\rm MoO_{3}^{s,l}}$, ${\rm MoO_{2.750}^{s}}$,
${\rm MoO_{2.875}^{s}}$, ${\rm MoO_{2.889}^{s}}$, ${\rm MoS_{2}^{s}}$,
${\rm Mo_{2}S_{3}^{s,l}}$, ${\rm MoCl_{4}^{s,l}}$, ${\rm MoCl_{5}^{s,l}}$,
${\rm MoC^{s}}$, ${\rm Mo^{g}}$, ${\rm MoO^{g}}$, ${\rm MoO_{2}^{g}}$,
${\rm MoO_{3}^{g}}$, ${\rm Mo_{2}O_{6}^{g}}$, ${\rm Mo_{3}O_{9}^{g}}$,
${\rm Mo_{4}O_{12}^{g}}$, ${\rm Mo_{5}O_{15}^{g}}$, ${\rm MoO_{2}Cl_{2}^{g}}$,
${\rm MoOCl^{g}}$,  ${\rm MoOCl_{2}^{g}}$, ${\rm MoOCl_{3}^{g}}$,
${\rm MoOCl_{4}^{g}}$, ${\rm MoO_{2}Cl^{g}}$, ${\rm MoCl^{g}}$,
${\rm MoCl_{2}^{g}}$, ${\rm MoCl_{3}^{g}}$, ${\rm MoCl_{4}^{g}}$,
${\rm MoCl_{5}^{g}}$, ${\rm MoCl_{6}^{g}}$, ${\rm Mo_{2}Cl_{10}^{g}}$,
${\rm Mo_{3}Cl_{15}^{g}}$, ${\rm MoH_{2}O_{4}^{g}}$, ${\rm MoF^{g}}$,
${\rm MoF_{2}^{g}}$, ${\rm MoF_{3}^{g}}$, ${\rm MoF_{4}^{g}}$,
${\rm MoF_{5}^{g}}$, ${\rm MoF_{6}^{g}}$, ${\rm Mo_{2}F_{10}^{g}}$,
${\rm Mo_{3}F_{15}^{g}}$ 

\paragraph*{Tungsten}

${\rm W^{s}}$, ${\rm WCl_{2}^{s}}$, ${\rm WCl_{4}^{s}}$, ${\rm WCl_{5}^{s,l}}$,
${\rm WCl_{6}^{l}}$, ${\rm WO_{2}^{s}}$, ${\rm WO_{3}^{s}}$, ${\rm WO_{2.72}^{s}}$,
${\rm WO_{2.9}^{s}}$, ${\rm WO_{2.96}^{s}}$, ${\rm WOCl_{4}^{l}}$,
${\rm WO_{2}Cl_{2}^{s}}$, ${\rm WS_{2}^{s}}$, ${\rm WH_{2}O_{4}^{s}}$,
${\rm WC^{s}}$, ${\rm W_{2}C^{s}}$, ${\rm WF_{6}^{l}},$ ${\rm W^{g}}$,
${\rm WCl^{g}}$, ${\rm WCl_{2}^{g}}$, ${\rm WCl_{4}^{g}}$, ${\rm WCl_{5}^{g}}$,
${\rm WCl_{6}^{g}}$, ${\rm W_{2}Cl_{10}^{g}}$, ${\rm WO^{g}}$,
${\rm WO_{2}^{g}}$, ${\rm WO_{3}^{g}}$, $({\rm WO_{3})_{2}^{g}}$,
$({\rm WO_{3})_{3}^{g}}$, $({\rm WO_{3})_{4}^{g}}$, $({\rm WO_{3})_{5}^{g}}$,
${\rm W_{3}O_{8}^{g}}$, ${\rm WO^{g}}$, ${\rm WO^{g}}$, ${\rm WOCl_{4}^{g}}$,
${\rm WO_{2}Cl_{2}^{g}}$, ${\rm WH_{2}O_{4}^{g}}$, ${\rm WF^{g}}$,
${\rm WF_{2}^{g}}$, ${\rm WF_{3}^{g}}$, ${\rm WF_{4}^{g}}$, ${\rm WF_{5}^{g}}$,
${\rm WF_{6}^{g}}$ 

\paragraph*{Sulfur}

${\rm S^{l,g}}$, ${\rm S_{2}^{g}}$, ${\rm S_{3}^{g}}$, ${\rm S_{4}^{g}}$,
${\rm S_{5}^{g}}$, ${\rm S_{6}^{g}}$, ${\rm S_{7}^{g}}$, ${\rm S_{8}^{g}}$,
${\rm SO_{2}^{g}}$, ${\rm SO_{3}^{g}}$, ${\rm S_{2}O^{g}}$, ${\rm SO^{g}}$,
${\rm SCl^{g}}$, ${\rm SCl_{2}^{g}},{\rm S_{2}Cl^{g}}$, ${\rm S_{2}Cl_{2}^{g}}$,
${\rm SOCl_{2}^{g}}$, ${\rm SO_{2}Cl_{2}^{g}}$, ${\rm SH^{g}}$,
${\rm H_{2}S^{g}}$, ${\rm H_{2}S_{2}^{g}}$, ${\rm H_{2}SO_{4}^{g}}$,
${\rm COS^{g}}$, ${\rm CS^{g}}$, ${\rm CS_{2}^{g}}$, ${\rm SF^{g}}$,
${\rm SF_{2}^{g}}$, ${\rm SF_{3}^{g}}$, ${\rm SF_{4}^{g}}$, ${\rm SF_{5}^{g}}$,
${\rm SF_{6}^{g}}$, ${\rm S_{2}F_{2}^{g}}$, ${\rm S_{2}F_{10}^{g}}$

\paragraph*{Chlorine and fluorine}

${\rm Cl^{g}}$, ${\rm Cl_{2}^{g}}$, ${\rm ClO^{g}}$, ${\rm Cl_{2}O^{g}}$,
${\rm ClO_{2}^{g}}$, ${\rm HCl^{g}}$, ${\rm F^{g}}$, ${\rm F_{2}^{g}}$,
${\rm HF^{g}}$, ${\rm (HF)_{2}^{g}}$, ${\rm (HF)_{3}^{g}}$, ${\rm (HF)_{4}^{g}}$,
${\rm (HF)_{5}^{g}}$, ${\rm (HF)_{6}^{g}}$, ${\rm (HF)_{7}^{g}}$

\paragraph*{Carbon}

${\rm C^{s,g}}$, ${\rm CH^{g}}$, ${\rm CH_{2}^{g}}$, ${\rm CH_{3}^{g}}$,
${\rm CH_{4}^{g}}$, ${\rm HCO^{g}}$, ${\rm CH_{2}O^{g}}$, ${\rm CH_{2}O_{2}^{g}}$,
${\rm CO^{g}}$, ${\rm CO_{2}^{g}}$,

\paragraph*{Oxygen and hydrogen}

${\rm O^{g}}$, ${\rm O_{2}^{g}}$, ${\rm O_{3}^{g}}$, ${\rm H^{g}}$,
${\rm H_{2}^{g}}$, ${\rm H_{2}O^{g}}$, ${\rm H_{2}O_{2}^{g}}$,
${\rm HO_{2}^{g}}$, ${\rm HO^{g}}$.

\subsection{Numerical example of calculating the metal precursor desorption flux for  CVD from metal oxide and elemental sulfur precursors.
  \label{app:calcul} }

In this section, we will show how to choose the parameters of the deposition process to ensure lateral gas-phase diffusion based on equilibrium calculations.

As an illustration,  we provide numerical examples of the calculation of the equilibrium desorption rate for the case of MoS\textsubscript{2} deposition using molybdenum oxide and elemental sulfur precursors.

\subsubsection*{Example 1: Metal precursor desorption rate in CVD using MoO\protect\textsubscript{3}
and S\protect\textsubscript{2 } precursors}

\begin{table}
\begin{tabular}{|c|c|c|c|c|c|c|}
\hline 
\multirow{2}{*}{T,\textcelsius} & \multicolumn{4}{c|}{$-G$, kJ/mol} & \multicolumn{2}{c|}{$\genfrac{}{}{0pt}{4}{\rm MoO_3^g+\frac{7}{4}S_2^g\rightarrow}{\rm \rightarrow MoS_2^s+\frac{3}{2}SO_2^g} $}\tabularnewline
\cline{2-7} \cline{3-7} \cline{4-7} \cline{5-7} \cline{6-7} \cline{7-7} 
 & $\rm MoO_3^g$ &$\rm S_2^g$ & $\rm MoS_2^s$ & $\rm SO_2$ & $-\Delta G_{r}\rm\frac{kJ}{mol}$  & K\tabularnewline
\hline 
\hline 
400 & 562 & 31 & 330 & 471 & 421 & $4.84\cdot10^{32}$\tabularnewline
\hline 
500 & 596 & 57 & 343 & 500 & 398 & $7.63\cdot10^{26}$\tabularnewline
\hline 
600 & 631 & 83 & 356 & 530 & 375 & $2.63\cdot10^{22}$\tabularnewline
\hline 
700 & 666 & 110 & 370 & 560 & 352 & $7.61\cdot10^{18}$\tabularnewline
\hline 
800 & 703 & 137 & 386 & 591 & 329 & $1.02\cdot10^{16}$\tabularnewline
\hline 
900 & 740 & 165 & 402 & 622 & 306 & $4.31\cdot10^{13}$\tabularnewline
\hline 
1000 & 778 & 192 & 418 & 653 & 284 & $4.35\cdot10^{11}$\tabularnewline
\hline 
\end{tabular}\caption{Gibbs free energies of main species involved in MoS\protect\textsubscript{2} CVD from $\rm MoO_3$ and $\rm S_2$  precursors.}
\label{tab:oxide-thermo}
\end{table}

Let's consider a process with the following parameters:
\begin{itemize}
\item Process pressure is 1 bar (100kPa), process temperature is 900\textcelsius{},
carrier gas is an inert gas.
\item Initial precursor partial pressures are $P_{{\rm MoO_{3}}}^{0}=10^{-6}$~atm,
$P_{{\rm S_{2}}}^{0}=10^{-3}$~atm, $P_{{\rm SO_{2}}}^{0}$=0~atm
\end{itemize}
For simplicity, we assume that the deposition process involves only 4 species:
${\rm MoO_{3}^{gas}}$, ${\rm S_{2}^{gas}}$, ${\rm MoS_{2}^{solid}}$,
and ${\rm SO_{2}^{gas}}$, connected via single deposition reaction
(Equation~\ref{eq:MoO-sulfurization}):

\begin{equation}
{\rm MoO_{3}^{gas}+\frac{7}{4}S_{2}^{gas}\underset{x}{\rightarrow}MoS_{2}^{solid}+\frac{3}{2}SO_{2}^{gas}}\label{eq:MoO-sulfurization-1}
\end{equation}
 where reaction quotient $x$ is limited by the amount of precursor
present in the deficiency: 
\begin{equation}
x\subset0..min(P_{{\rm MoO_{3}}}^{0},\frac{4}{7}P_{{\rm S_{2}}}^{0})
\end{equation}

The thermochemical properties of the species involved in this reaction are given in Table~\ref{tab:oxide-thermo}.

The equilibrium partial pressures of ${\rm MoO_{3}^{gas}}$, ${\rm S_{2}^{gas}}$, and ${\rm SO_{2}^{gas}}$ can be found by solving equation 

\begin{multline}
\frac{P_{{\rm SO_{2}}}^{eq^{\frac{3}{2}}}}{P_{{\rm S_{2}}}^{eq^{\frac{7}{4}}}P_{{\rm MoO_{3}}}^{eq}}=\frac{(\frac{3}{2}x^{eq})^{\frac{3}{2}}}{(P_{{\rm S_{2}}}^{0}-\frac{7}{2}x^{eq})^{\frac{7}{4}}(P_{{\rm MoO_{3}}}^{0}-x^{eq})}=\\
=K(T=900^{\circ}{\rm C})\approx4.31\cdot10^{13}\label{eq:MoS2-CVT-K}
\end{multline}

against $x^{eq}$. However, for this particular case, it can be simplified by
noting that the equilibrium is shifted to the right, so that $P_{{\rm MoO_{3}}}^{eq}\ll P_{{\rm MoO_{3}}}^{0}$,
and that $P_{{\rm S_{2}}}^{0}\gg P_{{\rm MoO_{3}}}^{0}$. Under this
assumption

\begin{equation}
\begin{cases}
P_{{\rm MoO_{3}}}^{eq}=P_{{\rm MoO_{3}}}^{0}-x^{eq}\ll P_{{\rm MoO_{3}}}^{0} & \Rightarrow x^{eq}\approx P_{{\rm MoO_{3}}}^{0}\\
P_{{\rm S_{2}}}^{eq}=P_{{\rm S_{2}}}^{0}-\frac{7}{4}x^{eq}\approx P_{{\rm S_{2}}}^{0}\\
P_{{\rm SO_{2}}}^{eq}=\frac{3}{2}x^{eq}\approx\frac{3}{2}P_{{\rm MoO_{3}}}^{0}
\end{cases}
\end{equation}
so that Equation~\ref{eq:MoS2-CVT-K} is simplified to:
\begin{multline}
\frac{P_{{\rm SO_{2}}}^{eq^{\frac{3}{2}}}}{P_{{\rm S_{2}}}^{eq^{\frac{7}{4}}}P_{{\rm MoO_{3}}}^{eq}}\approx\frac{(\frac{3}{2}P_{{\rm MoO_{3}}}^{0})^{\frac{3}{2}}}{(P_{{\rm S_{2}}}^{0}){}^{\frac{7}{4}}P_{{\rm MoO_{3}}}^{eq}}=\\
=K(T=900^{\circ}{\rm C})\approx4.31\cdot10^{13}\label{eq:MoO3-S2-simpified}
\end{multline}

Substituting numerical values of $P_{{\rm MoO_{3}}}^{0}$ and $P_{{\rm S_{2}}}^{0}$
into Equation~\ref{eq:MoO3-S2-simpified} gives:

$$
P_{{\rm MoO_{3}}}^{eq}(T=900^{\circ}{\rm C})\approx7.6\times10^{-18}[{\rm atm}]
$$

Substituting this value in Equation~\ref{eq:des_flux_eq} gives equilibrium desorption flux:
\begin{multline}
J_{des}^{eq}=\frac{P_{{\rm MoO_{3}}}^{eq}}{\sqrt{2\pi m_{{\rm MoO_{3}}}k_{B}T}}=\\
=\frac{7.6\times10^{-13}[{\rm Pa}]}{\sqrt{6.28\cdot2.39\times10^{-25}[{\rm kg}]\cdot1.38\times10^{-23}[\rm{J/K}]\cdot1173.15[{\rm K}]}}=\\
=4.9\times10^{9}\left[\frac{1}{\rm m^{2}s}\right]\approx4.9\times10^{-10}\left[\frac{{\rm monolayer}}{{\rm s}}\right]\label{eq:des_flux_eq-MoO3}
\end{multline}
that is negligibly small, so that criterion \ref{eq:criterion} can be satisfied only for deposition rate slower than 1 monolayer per century.

\subsubsection*{Example 2: Increasing desorption rate by co-injecting reaction products}

The equilibrium pressure of molybdenum oxide $P_{{\rm MoO_{3}}}^{eq}$ can be increased by increasing concentration of ${\rm SO_{2}}$ in the gas mixture. Consider an increase in the equilibrium pressure
of ${\rm MO_{3}}$ in the case where the concentration of ${\rm SO_{2}}$
in the gas mixture is 10\%. The initial precursor partial pressures are $P_{{\rm MoO_{3}}}^{0}=10^{-6}$~atm, $P_{{\rm S_{2}}}^{0}=10^{-3}$~atm,
$P_{{\rm SO_{2}}}^{0}$=0.1~atm.

Assuming that

$$
\begin{cases}
P_{{\rm MoO_{3}}}^{eq}=P_{{\rm MoO_{3}}}^{0}-x^{eq}\ll P_{{\rm MoO_{3}}}^{0} & \Rightarrow x^{eq}\approx P_{{\rm MoO_{3}}}^{0}\\
P_{{\rm S_{2}}}^{eq}=P_{{\rm S_{2}}}^{0}-\frac{7}{4}x^{eq}\approx P_{{\rm S_{2}}}^{0}\\
P_{{\rm SO_{2}}}^{eq}=P_{{\rm SO_{2}}}^{0}+\frac{3}{2}x^{eq}\approx P_{{\rm SO_{2}}}^{0}
\end{cases}
$$

Equation~\ref{eq:MoS2-CVT-K} is approximated as:

\begin{equation}
\frac{P_{{\rm SO_{2}}}^{eq^{\frac{3}{2}}}}{P_{{\rm S_{2}}}^{eq^{\frac{7}{4}}}P_{{\rm MoO_{3}}}^{eq}}\approx\frac{(P_{{\rm SO_{2}}}^{0})^{\frac{3}{2}}}{(P_{{\rm S_{2}}}^{0}){}^{\frac{7}{4}}P_{{\rm MoO_{3}}}^{eq}}=K(T=900^{\circ}{\rm C})\approx4.31\cdot10^{13}
\end{equation}

that gives significantly higher equilibrium vapor pressure and desorption
rate:

$$
P_{{\rm MoO_{3}}}^{eq}(T=900^{\circ}{\rm C})\approx1.3\times10^{-10}[{\rm atm}]
$$

\begin{multline}
J_{des}^{eq}=\frac{P_{{\rm MoO_{3}}}^{eq}}{\sqrt{2\pi m_{{\rm MoO_{3}}}k_{B}T}}=\\
=\frac{1.3\times10^{-5}[{\rm Pa}]}{\sqrt{6.28\cdot2.39\times10^{-25}[{\rm kg}]\cdot1.38\times10^{-23}[\rm{J/K}]\cdot1173.15[{\rm K}]}}=\\
=8.4\times10^{16}\left[\frac{1}{\rm{m^{2}s}}\right]\approx1/120\left[\frac{{\rm monolayer}}{{\rm s}}\right]\label{eq:des_flux_eq-MoO3-SO2}
\end{multline}

so condition \ref{eq:criterion} is satisfied for deposition rate
$J_{net}\ll$ 1 monolayer per 2 minutes.

\subsubsection*{Example 3: Multicomponent equilibrium analysis}

\begin{table}
\begin{tabular}{|>{\centering}m{0.22\columnwidth}|>{\centering}p{0.65\columnwidth}|}
\hline 
Elements & Substances\tabularnewline
\hline 
\hline 
Molybdenum & ${\rm Mo^{s,g}}$\tabularnewline
\hline 
Sulfur & ${\rm S^{l,g}}$, ${\rm S_{2}^{g}}$, ${\rm S_{3}^{g}}$, ${\rm S_{4}^{g}}$,
${\rm S_{5}^{g}}$, ${\rm S_{6}^{g}}$, ${\rm S_{7}^{g}}$, ${\rm S_{8}^{g}}$\tabularnewline
\hline 
Oxygen & ${\rm O^{g}}$, ${\rm O_{2}^{g}}$, ${\rm O_{3}^{g}}$\tabularnewline
\hline 
Molybdenum oxides & ${\rm MoO^{g}}$, ${\rm MoO_{2}^{s,g}}$, ${\rm MoO_{3}^{s,l,g}}$,

${\rm Mo_{2}O_{6}^{g}}$, ${\rm Mo_{3}O_{9}^{g}}$, ${\rm Mo_{4}O_{12}^{g}}$,

${\rm Mo_{5}O_{15}^{g}}$, ${\rm MoO_{2.750}^{s}}$, ${\rm MoO_{2.875}^{s}}$,
${\rm MoO_{2.889}^{s}}$ \tabularnewline
\hline 
Molybdenum sulfides & ${\rm MoS_{2}^{s}}$, ${\rm Mo_{2}S_{3}^{s,l}}$ \tabularnewline
\hline 
Sulfur oxides & ${\rm SO_{2}^{g}}$, ${\rm SO_{3}^{g}}$, ${\rm S_{2}O^{g}}$, ${\rm SO^{g}}$\tabularnewline
\hline 
\end{tabular}\caption{\label{tab:MoS2-CVT-subs}Substances considered in equilibrium calculation of $\rm MoS_2$ deposition by CVD from $\rm MoO_3$ and $\rm S_2$ precursors. The aggregate state of substance is indicated by a superscript: s -- solid, l -- liquid, g -- gas.}

\end{table}

The accuracy of thermodynamic equilibrium analysis depends on whether
all significant substances are included in the calculation.
Simplified calculations involving only one basic chemical reaction, such as those discussed above, are intuitive, but if the substances involved in them are chosen incorrectly, they will give erroneous results.

Numerical calculation of the equilibrium composition of the multicomponent MoS\textsubscript{2} CVD from $\rm MoO_3$ and $\rm S_2$ precursors was performed by Gibbs energy minimization. All substances listed in the  Table~\ref{tab:MoS2-CVT-subs} are considered. The resulting values of the equilibrium pressures of the components are shown in Figure~\ref{fig:numeric-example}b,d.

As can be seen from  Figures~\ref{fig:numeric-example}a and  \ref{fig:numeric-example}b, the results obtained in the single-reaction approximation (Equation~\ref{eq:des_flux_eq-MoO3}) taking into account only one gaseous molybdenum compound - $\rm MoO_3$   are close to the results obtained by the multicomponent calculation, since the equilibrium pressure of other gaseous molybdenum compounds is negligible compared to the equilibrium pressure of $\rm MoO_3$.

At the same time, the results of the calculation of the system with the addition of $\rm SO_2$, shown in Figure~\ref{fig:numeric-example}c and Figure~\ref{fig:numeric-example}d, show significant differences compared to the single-reaction approximation:
\begin{enumerate}
  \item  equilibrium desorption rate is several times higher than predicted by the simplified calculation using the Equation~\ref{eq:des_flux_eq-MoO3-SO2}, since, in addition to $\rm MoO_3$, other molecules, including $\rm Mo_2O_6$ and $\rm Mo_3O_9$, are present in comparable amounts.
\item  equilibrium composition of the condensed phase changes at 917\textcelsius{}: $\rm MoS_2$ is the equilibrium condensed molybdenum compound at temperatures below 917\textcelsius{},  while  $\rm MoO_2$ is formed in equilibrium  at temperatures above 917\textcelsius{}. 
  \end{enumerate}

This emphasizes the importance of multicomponent analysis, since ignoring molecules can lead to a
significant distortion of the results.

\subsection{Extending the model to 2D TMD annealing}
A significant part of the literature is devoted to the synthesis of two-dimensional TMD crystals by chalcogenation of pre-deposited thin films of a transition metal \cite{shahzad2017effects,yim2018wide} or its compounds \cite{chiappe2016controlled,giannazzo2023atomic,zhang2020universal,esposito2023role}.
In contrast to CVD processes, the condensed phase composition changes continuously during chalcogenation, so that the surface remains in non-equilibrium conditions until chalcogenation is completed.
The model proposed in this paper assumes a near-equilibrium state on the surface and thus cannot be directly used to analyze chalcogenation.

However, the model can be extended to assess the TMD layer annealing under stationary conditions by the changes briefly outlined below.

First, efficient lateral gas-phase transport relates to the possibility of metal desorption from the starting surface during the annealing process.
Let us consider the annealing of a TMDC film with a surface atomic density $N_s$ carried out for a time $\tau_{a}$.
The total number of metal atoms that have undergone desorption $N_{des}$ during the process is
\begin{equation}
  N_{des} = J_{des}\tau_{a}
\end{equation}
The gas-phase diffusion contributes significantly to the lateral transport when
\begin{equation}
N_{des}\gg N_s
\end{equation}
Second, the desorption flux can be estimated using the equilibrium desorption flux calculated by Equation \eqref{eq:des_flux_eq} as an upper boundary: 
\begin{equation}
  \underset{{\rm Slow}\xleftarrow[{\rm }]{{\rm surface\,kinetics}}{\rm Fast}}{0 < J_{des}^{kin} < J_{des}^{eq}}\label{eq:J-interval-anneal}
\end{equation}

Third, when annealing in an open-flow system, some amount of material is lost due to evaporation through the boundary layer. 
The evaporation rate $J_{evap}$ is proportional to the partial pressure of volatile metal species at the substrate surface:
\begin{equation}
J_{evap} = D_{Me}\frac{P_{Me}^{surf}}{\delta}\label{eq:diff-lim-anneal}
\end{equation}
and reaches its greatest value $J_{evap}^{eq}$ when the surface reaction is close to equilibrium:
\begin{equation}
  \underset{{\rm Slow}\xleftarrow[{\rm }]{{\rm surface\,kinetics}}{\rm Fast}}{0 \le J_{evap} \le J_{evap}^{eq}\equiv  D_{Me}\frac{P_{Me}^{eq}}{\delta} }\label{eq:Jevap-interval-anneal}
\end{equation}
Thus, the equilibrium calculation allows to estimate the upper limit of material loss $N_{evap}$ during the annealing process:
\begin{equation}
N_{evap}=J_{evap}\tau_{a} \le J_{evap}^{eq} \tau_{a} 
\end{equation}
When the condition
\begin{equation}
  \label{eq:loss-neglig}
   J_{evap}^{eq} \tau_{a}  \ll N_s
\end{equation}
is met, the possible material loss due to evaporation is negligible.

\section{Acknowledgments}

The authors gratefully acknowledge the funding from the imec IIAP
core CMOS programs and the European Union’s Graphene Flagship
grant agreement No~952792.

\bibliographystyle{unsrt}
\bibliography{mos2-review}

\begin{thebibliography}{10}

\bibitem{robinson2018perspective}
Joshua~A Robinson.
\newblock Perspective: 2d for beyond cmos.
\newblock {\em Apl Materials}, 6(5):058202, 2018.

\bibitem{wang2021road}
Shuiyuan Wang, Xiaoxian Liu, and Peng Zhou.
\newblock The road for 2d semiconductors in the silicon age.
\newblock {\em Advanced Materials}, page 2106886, 2021.

\bibitem{schmidt2015electronic}
Hennrik Schmidt, Francesco Giustiniano, and Goki Eda.
\newblock Electronic transport properties of transition metal dichalcogenide
  field-effect devices: surface and interface effects.
\newblock {\em Chemical Society Reviews}, 44(21):7715--7736, 2015.

\bibitem{li2016charge}
Song-Lin Li, Kazuhito Tsukagoshi, Emanuele Orgiu, and Paolo Samor{\`\i}.
\newblock Charge transport and mobility engineering in two-dimensional
  transition metal chalcogenide semiconductors.
\newblock {\em Chemical Society Reviews}, 45(1):118--151, 2016.

\bibitem{manzeli20172d}
Sajedeh Manzeli, Dmitry Ovchinnikov, Diego Pasquier, Oleg~V Yazyev, and Andras
  Kis.
\newblock 2d transition metal dichalcogenides.
\newblock {\em Nature Reviews Materials}, 2(8):1--15, 2017.

\bibitem{yang2022oxidations}
Junqiang Yang, Xiaochi Liu, Qianli Dong, Yaqi Shen, Yuchuan Pan, Zhongwang
  Wang, Kui Tang, Xianfu Dai, Rongqi Wu, Yuanyuan Jin, et~al.
\newblock Oxidations of two-dimensional semiconductors: Fundamentals and
  applications.
\newblock {\em Chinese Chemical Letters}, 33(1):177--185, 2022.

\bibitem{Radisavljevic2011}
Branimir Radisavljevic, Aleksandra Radenovic, Jacopo Brivio, Valentina
  Giacometti, and Andras Kis.
\newblock Single-layer mos2 transistors.
\newblock {\em Nature nanotechnology}, 6(3):147--150, 2011.

\bibitem{Georgiou2013}
Thanasis Georgiou, Rashid Jalil, Branson~D Belle, Liam Britnell, Roman~V
  Gorbachev, Sergey~V Morozov, Yong-Jin Kim, Ali Gholinia, Sarah~J Haigh, Oleg
  Makarovsky, et~al.
\newblock Vertical field-effect transistor based on graphene--ws2
  heterostructures for flexible and transparent electronics.
\newblock {\em Nature nanotechnology}, 8(2):100--103, 2013.

\bibitem{li2020large}
Na~Li, Qinqin Wang, Cheng Shen, Zheng Wei, Hua Yu, Jing Zhao, Xiaobo Lu, Guole
  Wang, Congli He, Li~Xie, et~al.
\newblock Large-scale flexible and transparent electronics based on monolayer
  molybdenum disulfide field-effect transistors.
\newblock {\em Nature Electronics}, 3(11):711--717, 2020.

\bibitem{Ly2016grain-boundary-scattering}
Thuc~Hue Ly, David~J Perello, Jiong Zhao, Qingming Deng, Hyun Kim, Gang~Hee
  Han, Sang~Hoon Chae, Hye~Yun Jeong, and Young~Hee Lee.
\newblock Misorientation-angle-dependent electrical transport across molybdenum
  disulfide grain boundaries.
\newblock {\em Nature communications}, 7(1):1--7, 2016.

\bibitem{Kim2017grain-boundary-mobility}
Jae-Keun Kim, Younggul Song, Tae-Young Kim, Kyungjune Cho, Jinsu Pak,
  Barbara~Yuri Choi, Jiwon Shin, Seungjun Chung, and Takhee Lee.
\newblock Analysis of noise generation and electric conduction at grain
  boundaries in cvd-grown mos2 field effect transistors.
\newblock {\em Nanotechnology}, 28(47):47LT01, 2017.

\bibitem{routledge1970nucleation}
KJ~Routledge and MJ~Stowell.
\newblock Nucleation kinetics in thin film growth. i. computer simulation of
  nucleation and growth behaviour.
\newblock {\em Thin Solid Films}, 6(6):407--421, 1970.

\bibitem{peters2020directing}
Lisanne Peters, Cormac {\'O}~Coile{\'a}in, Patryk Dluzynski, Rita Siris,
  Georg~S Duesberg, and Niall McEvoy.
\newblock Directing the morphology of chemical vapor deposition-grown mos2 on
  sapphire by crystal plane selection.
\newblock {\em physica status solidi (a)}, 217(15):2000073, 2020.

\bibitem{Kim2017}
HoKwon Kim, Dmitry Ovchinnikov, Davide Deiana, Dmitrii Unuchek, and Andras Kis.
\newblock Suppressing nucleation in metal--organic chemical vapor deposition of
  mos2 monolayers by alkali metal halides.
\newblock {\em Nano letters}, 17(8):5056--5063, 2017.

\bibitem{choudhury2018}
Tanushree~H Choudhury, Hamed Simchi, Raphael Boichot, Mikhail Chubarov,
  Suzanne~E Mohney, and Joan~M Redwing.
\newblock Chalcogen precursor effect on cold-wall gas-source chemical vapor
  deposition growth of ws2.
\newblock {\em Crystal Growth \& Design}, 18(8):4357--4364, 2018.

\bibitem{Park2015}
Jusang Park, Wonseon Lee, Taejin Choi, Sung-Hwan Hwang, Jae~Min Myoung,
  Jae-Hoon Jung, Soo-Hyun Kim, and Hyungjun Kim.
\newblock Layer-modulated synthesis of uniform tungsten disulfide nanosheet
  using gas-phase precursors.
\newblock {\em Nanoscale}, 7(4):1308--1313, 2015.

\bibitem{Okada2021-WF6-H2S}
Mitsuhiro Okada, Naoya Okada, Wen-Hsin Chang, Tetsuo Shimizu, Toshitaka Kubo,
  Masatou Ishihara, and Toshifumi Irisawa.
\newblock Micrometer-scale ws2 atomic layers grown by alkali metal free
  gas-source chemical vapor deposition with h2s and wf6 precursors.
\newblock {\em Japanese Journal of Applied Physics}, 60(SB):SBBH09, 2021.

\bibitem{Chen2015oxyden}
Wei Chen, Jing Zhao, Jing Zhang, Lin Gu, Zhenzhong Yang, Xiaomin Li, Hua Yu,
  Xuetao Zhu, Rong Yang, Dongxia Shi, Xuechun Lin, Jiandong Guo, Xuedong Bai,
  and Guangyu Zhang.
\newblock Oxygen-assisted chemical vapor deposition growth of large
  single-crystal and high-quality monolayer mos2.
\newblock {\em Journal of the American Chemical Society}, 137(50):15632--15635,
  2015.
\newblock PMID: 26623946.

\bibitem{Yu2013}
Yifei Yu, Chun Li, Yi~Liu, Liqin Su, Yong Zhang, and Linyou Cao.
\newblock Controlled scalable synthesis of uniform, high-quality monolayer and
  few-layer mos2 films.
\newblock {\em Scientific reports}, 3(1):1--6, 2013.

\bibitem{Okada2014}
Mitsuhiro Okada, Takumi Sawazaki, Kenji Watanabe, Takashi Taniguch, Hiroki
  Hibino, Hisanori Shinohara, and Ryo Kitaura.
\newblock Direct chemical vapor deposition growth of ws2 atomic layers on
  hexagonal boron nitride.
\newblock {\em ACS nano}, 8(8):8273--8277, 2014.

\bibitem{Hu2017}
Dake Hu, Guanchen Xu, Lei Xing, Xingxu Yan, Jingyi Wang, Jingying Zheng,
  Zhixing Lu, Peng Wang, Xiaoqing Pan, and Liying Jiao.
\newblock Two-dimensional semiconductors grown by chemical vapor transport.
\newblock {\em Angewandte Chemie International Edition}, 56(13):3611--3615,
  2017.

\bibitem{Lan2020}
Feifei Lan, Ruixia Yang, Song Hao, Baozeng Zhou, Kewei Sun, Hongjuan Cheng,
  Song Zhang, Lujie Li, and Lei Jin.
\newblock Controllable synthesis of millimeter-size single crystal ws2.
\newblock {\em Applied Surface Science}, 504:144378, 2020.

\bibitem{Chung1998}
J.-W. Chung, Z.R. Dai, and F.S. Ohuchi.
\newblock {WS$_2$ thin films by metal organic chemical vapor deposition}.
\newblock {\em Journal of Crystal Growth}, 186(1):137--150, mar 1998.

\bibitem{Carmalt2003}
Claire~J Carmalt, Ivan~P Parkin, and Emily~S Peters.
\newblock Atmospheric pressure chemical vapour deposition of ws2 thin films on
  glass.
\newblock {\em Polyhedron}, 22(11):1499--1505, 2003.

\bibitem{mun2016low}
Jihun Mun, Yeongseok Kim, Il-Suk Kang, Sung~Kyu Lim, Sang~Jun Lee, Jeong~Won
  Kim, Hyun~Min Park, Taesung Kim, and Sang-Woo Kang.
\newblock Low-temperature growth of layered molybdenum disulphide with
  controlled clusters.
\newblock {\em Scientific reports}, 6(1):21854, 2016.

\bibitem{Cadot2017}
Stéphane Cadot, Olivier Renault, Denis Rouchon, Denis Mariolle, Emmanuel
  Nolot, Chloé Thieuleux, Laurent Veyre, Hanako Okuno, François Martin, and
  Elsje~Alessandra Quadrelli.
\newblock Low-temperature and scalable cvd route to ws2 monolayers on sio2/si
  substrates.
\newblock {\em Journal of Vacuum Science \& Technology A}, 35(6):061502, 2017.

\bibitem{kim2017wafer}
TaeWan Kim, Jihun Mun, Hyeji Park, DaeHwa Joung, Mangesh Diware, Chegal Won,
  Jonghoo Park, Soo-Hwan Jeong, and Sang-Woo Kang.
\newblock Wafer-scale production of highly uniform two-dimensional mos2 by
  metal-organic chemical vapor deposition.
\newblock {\em Nanotechnology}, 28(18):18LT01, 2017.

\bibitem{chiappe2018layer}
Daniele Chiappe, Jonathan Ludwig, Alessandra Leonhardt, Salim El~Kazzi,
  Ankit~Nalin Mehta, Thomas Nuytten, Umberto Celano, Surajit Sutar, Geoffrey
  Pourtois, Matty Caymax, et~al.
\newblock Layer-controlled epitaxy of 2d semiconductors: bridging nanoscale
  phenomena to wafer-scale uniformity.
\newblock {\em Nanotechnology}, 29(42):425602, 2018.

\bibitem{cwik2018direct}
Stefan Cwik, Dariusz Mitoraj, Oliver Mendoza~Reyes, Detlef Rogalla, Daniel
  Peeters, Jiyeon Kim, Hanno~Maria Sch{\"u}tz, Claudia Bock, Radim Beranek, and
  Anjana Devi.
\newblock Direct growth of mos2 and ws2 layers by metal organic chemical vapor
  deposition.
\newblock {\em Advanced Materials Interfaces}, 5(16):1800140, 2018.

\bibitem{grundmann2019h2s}
A~Grundmann, D~Andrzejewski, T~K{\"u}mmell, G~Bacher, M~Heuken, H~Kalisch, and
  A~Vescan.
\newblock H 2 s-free metal-organic vapor phase epitaxy of coalesced 2d ws 2
  layers on sapphire.
\newblock {\em MRS Advances}, 4:593--599, 2019.

\bibitem{shinde2019rapid}
Nitin~Babu Shinde, Bellarmine Francis, MS~Ramachandra~Rao, Beo~Deul Ryu,
  S~Chandramohan, and Senthil~Kumar Eswaran.
\newblock Rapid wafer-scale fabrication with layer-by-layer thickness control
  of atomically thin mos2 films using gas-phase chemical vapor deposition.
\newblock {\em APL Materials}, 7(8):081113, 2019.

\bibitem{Cohen2020}
Assael Cohen, Avinash Patsha, Pranab~K Mohapatra, Miri Kazes, Kamalakannan
  Ranganathan, Lothar Houben, Dan Oron, and Ariel Ismach.
\newblock Growth-etch metal--organic chemical vapor deposition approach of ws2
  atomic layers.
\newblock {\em ACS nano}, 15(1):526--538, 2020.

\bibitem{seol2020high}
Minsu Seol, Min-Hyun Lee, Haeryong Kim, Keun~Wook Shin, Yeonchoo Cho, Insu
  Jeon, Myoungho Jeong, Hyung-Ik Lee, Jiwoong Park, and Hyeon-Jin Shin.
\newblock High-throughput growth of wafer-scale monolayer transition metal
  dichalcogenide via vertical ostwald ripening.
\newblock {\em Advanced Materials}, 32(42):2003542, 2020.

\bibitem{schaefer2021carbon}
Christian~M Schaefer, Jose~M Caicedo~Roque, Guillaume Sauthier, Jessica
  Bousquet, Clement Hebert, Justin~R Sperling, Amador Perez-Tomas, Jose
  Santiso, Elena del Corro, and Jose~A Garrido.
\newblock Carbon incorporation in mocvd of mos2 thin films grown from an
  organosulfide precursor.
\newblock {\em Chemistry of Materials}, 33(12):4474--4487, 2021.

\bibitem{Arctowski1895}
Henryk Arctowski.
\newblock {Uber die doppelte Umsetzung bei gasformigen Korpern}.
\newblock {\em Zeitschrift fur anorganische Chemie}, 8(1):213--223, 1895.

\bibitem{Imanishi1992}
Nobuyuki Imanishi, Kiyoshi Kanamura, and Zen ichiro Takehara.
\newblock {Synthesis of MoS$_2$ Thin Film by Chemical Vapor Deposition Method
  and Discharge Characteristics as a Cathode of the Lithium Secondary Battery}.
\newblock {\em Journal of The Electrochemical Society}, 139(8):2082--2087, aug
  1992.

\bibitem{Keune2000}
{Keune, H.}, {Lacom, W.}, {Rossi, F.}, {Stoffels, E.}, {Stoffels, W. W.}, and
  {Wahl, G.}
\newblock {Formation and deposition of MoS$_2$-nanoparticles}.
\newblock {\em J. Phys. IV France}, 10:Pr2--19--Pr2--26, 2000.

\bibitem{Margolin2008WClx+H2S}
A~Margolin, FL~Deepak, Ronit Popovitz-Biro, M~Bar-Sadan, Y~Feldman, and
  R~Tenne.
\newblock Fullerene-like ws2 nanoparticles and nanotubes by the vapor-phase
  synthesis of wcln and h2s.
\newblock {\em Nanotechnology}, 19(9):095601, 2008.

\bibitem{Huang2014}
Chung-Che Huang, Feras Al-Saab, Yudong Wang, Jun-Yu Ou, John~C Walker, Shuncai
  Wang, Behrad Gholipour, Robert~E Simpson, and Daniel~W Hewak.
\newblock Scalable high-mobility mos 2 thin films fabricated by an atmospheric
  pressure chemical vapor deposition process at ambient temperature.
\newblock {\em Nanoscale}, 6(21):12792--12797, 2014.

\bibitem{Campbell2022WCl6-H2S}
William~R Campbell, Francesco Reale, Ravi Sundaram, and Simon~J Bending.
\newblock Optimisation of processing conditions during cvd growth of 2d ws2
  films from a chloride precursor.
\newblock {\em Journal of Materials Science}, pages 1--15, 2022.

\bibitem{Kim2022MeOCl4}
Jee~Hyeon Kim, Chaehyeon Ahn, Jong-Guk Ahn, Younghee Park, Soyoung Kim, Daehyun
  Kim, Jaeyoon Baik, Jaehoon Jung, and Hyunseob Lim.
\newblock Vapor pressure-controllable molecular inorganic precursors for growth
  of monolayer ws2: Influence of precursor-substrate interaction on growth
  thermodynamics.
\newblock {\em Applied Surface Science}, 587:152829, 2022.

\bibitem{lee1994preparation}
Woo~Y. Lee, Theodore~M. Besmann, and Michael~W. Stott.
\newblock Preparation of {MoS}2 thin films by chemical vapor deposition.
\newblock {\em Journal of Materials Research}, 9(6):1474--1483, jun 1994.

\bibitem{Groven2019}
B.~Groven, D.~Claes, A.~Nalin~Mehta, H.~Bender, W.~Vandervorst, M.~Heyns,
  M.~Caymax, I.~Radu, and A.~Delabie.
\newblock Chemical vapor deposition of monolayer-thin ws2 crystals from the wf6
  and h2s precursors at low deposition temperature.
\newblock {\em The Journal of Chemical Physics}, 150(10):104703, 2019.

\bibitem{Sahoo2018}
Prasana~K Sahoo, Shahriar Memaran, Yan Xin, Luis Balicas, and Humberto~R
  Guti{\'e}rrez.
\newblock One-pot growth of two-dimensional lateral heterostructures via
  sequential edge-epitaxy.
\newblock {\em Nature}, 553(7686):63--67, 2018.

\bibitem{Zhao2019}
Yuzhou Zhao and Song Jin.
\newblock Controllable water vapor assisted chemical vapor transport synthesis
  of ws2--mos2 heterostructure.
\newblock {\em ACS Materials Letters}, 2(1):42--48, 2019.

\bibitem{zhang2022additive-assisted}
Qing Zhang, Xixi Xiao, Lin Li, Dechao Geng, Wei Chen, and Wenping Hu.
\newblock Additive-assisted growth of scaled and quality 2d materials.
\newblock {\em Small}, 18(17):2107241, 2022.

\bibitem{Kang2015mo-decomposition}
Kibum Kang, Saien Xie, Lujie Huang, Yimo Han, Pinshane~Y Huang, Kin~Fai Mak,
  Cheol-Joo Kim, David Muller, and Jiwoong Park.
\newblock High-mobility three-atom-thick semiconducting films with wafer-scale
  homogeneity.
\newblock {\em Nature}, 520(7549):656--660, 2015.

\bibitem{Kastl2017}
Christoph Kastl, Christopher~T Chen, Tevye Kuykendall, Brian Shevitski,
  Thomas~P Darlington, Nicholas~J Borys, Andrey Krayev, P~James Schuck, Shaul
  Aloni, and Adam~M Schwartzberg.
\newblock The important role of water in growth of monolayer transition metal
  dichalcogenides.
\newblock {\em 2D Materials}, 4(2):021024, 2017.

\bibitem{Choi2017}
Soo~Ho Choi, Boandoh Stephen, Ji-Hoon Park, Joo~Song Lee, Soo~Min Kim, Woochul
  Yang, and Ki~Kang Kim.
\newblock Water-assisted synthesis of molybdenum disulfide film with single
  organic liquid precursor.
\newblock {\em Scientific reports}, 7(1):1--8, 2017.

\bibitem{pound1963condensation}
G~Pound and J~Hirth.
\newblock Condensation and evaporation, nucleation and growth kinetics.
\newblock {\em Prog. Mater. Sci}, 11, 1963.

\bibitem{bloem1977nucleation}
J~Bloem.
\newblock Nucleation in the epitaxial growth of silicon.
\newblock {\em Journal of Crystal Growth}, 38(3):364--366, 1977.

\bibitem{yamaguchi1993lateral}
Ko-ichi Yamaguchi Ko-ichi Yamaguchi and Kotaro Okamoto~Kotaro Okamoto.
\newblock Lateral supply mechanisms in selective metalorganic chemical vapor
  deposition.
\newblock {\em Japanese journal of applied physics}, 32(4R):1523, 1993.

\bibitem{coltrin2003mass}
Michael~E Coltrin and Christine~C Mitchell.
\newblock Mass transport and kinetic limitations in mocvd selective-area
  growth.
\newblock {\em Journal of crystal growth}, 254(1-2):35--45, 2003.

\bibitem{shioda2009selective}
Tomonari Shioda, Yuki Tomita, Masakazu Sugiyama, Yukihiro Shimogaki, and
  Yoshiaki Nakano.
\newblock Selective area metal--organic vapor phase epitaxy of nitride
  semiconductors for multicolor emission.
\newblock {\em IEEE Journal of Selected Topics in Quantum Electronics},
  15(4):1053--1065, 2009.

\bibitem{Gupta1999ELOG}
A~Gupta, XW~Li, S~Guha, and Gang Xiao.
\newblock Selective-area and lateral overgrowth of chromium dioxide (cro 2)
  films by chemical vapor deposition.
\newblock {\em Applied Physics Letters}, 75(19):2996--2998, 1999.

\bibitem{OLSSON200624}
Fredrik Olsson, Tiankai Zhu, Gaël Mion, and Sebastian Lourdudoss.
\newblock Large mask area effects in selective area growth.
\newblock {\em Journal of Crystal Growth}, 289(1):24--30, 2006.

\bibitem{Naumovets1985}
AG~Naumovets and Yu~S Vedula.
\newblock Surface diffusion of adsorbates.
\newblock {\em Surface Science Reports}, 4(7-8):365--434, 1985.

\bibitem{bradbury1984control}
DR~Bradbury, TI~Kamins, and C-W Tsao.
\newblock Control of lateral epitaxial chemical vapor deposition of silicon
  over insulators.
\newblock {\em Journal of applied physics}, 55(2):519--523, 1984.

\bibitem{tsuchiya1999cl}
Tomonobu Tsuchiya.
\newblock Cl-assisted selective area growth of inp by metalorganic vapor phase
  epitaxy.
\newblock {\em Japanese journal of applied physics}, 38(2S):1034, 1999.

\bibitem{Goodman1974}
DW~Shaw.
\newblock {\em Crystal Growth}, volume~1, chapter Mechanisms in Vapour Epitaxy
  of Semiconductors, pages 1--48.
\newblock Springer, 1974.

\bibitem{Yu2020vls}
Chenqi Yu, Pu~Chang, Lixiu Guan, and Junguang Tao.
\newblock Inward growth of monolayer mos2 single crystals from molten na2moo4
  droplets.
\newblock {\em Materials Chemistry and Physics}, 240:122203, 2020.

\bibitem{lee2012synthesis}
Yi-Hsien Lee, Xin-Quan Zhang, Wenjing Zhang, Mu-Tung Chang, Cheng-Te Lin,
  Kai-Di Chang, Ya-Chu Yu, Jacob Tse-Wei Wang, Chia-Seng Chang, Lain-Jong Li,
  et~al.
\newblock Synthesis of large-area mos2 atomic layers with chemical vapor
  deposition.
\newblock {\em Advanced materials}, 24(17):2320--2325, 2012.

\bibitem{Hyun2018MoO3-S2}
Cheol-Min Hyun, Jeong-Hun Choi, Seung~Won Lee, Jeong~Hwa Park, Kang-Taek Lee,
  and Ji-Hoon Ahn.
\newblock Synthesis mechanism of mos2 layered crystals by chemical vapor
  deposition using moo3 and sulfur powders.
\newblock {\em Journal of Alloys and Compounds}, 765:380--384, 2018.

\bibitem{Johnson1982moo3_NaCl}
DA~Johnson, JH~Levy, JC~Taylor, AB~Waugh, and J~Brough.
\newblock Purification of molybdenum: volatilisation processes using moo3.
\newblock {\em Polyhedron}, 1(5):479--482, 1982.

\bibitem{Belton1964}
GR~Belton and RL~McCarron.
\newblock The volatilization of tungsten in the presence of water vapor.
\newblock {\em The Journal of Physical Chemistry}, 68(7):1852--1856, 1964.

\bibitem{Belton1965}
GR~Belton and AS~Jordan.
\newblock The volatilization of molybdenum in the presence of water vapor.
\newblock {\em The Journal of Physical Chemistry}, 69(6):2065--2071, 1965.

\bibitem{Schaefer1973}
Harald Sch{\"a}fer, Theodor Grofe, and Marita Trenkel.
\newblock The chemical transport of molybdenum and tungsten and of their
  dioxides and sulfides.
\newblock {\em Journal of Solid State Chemistry}, 8(1):14--28, 1973.

\bibitem{Zheng2017MoO3-S2-separate}
Binjie Zheng and Yuanfu Chen.
\newblock Controllable growth of monolayer mos2 and mose2 crystals using
  three-temperature-zone furnace.
\newblock In {\em IOP conference series: materials science and engineering},
  volume 274, page 012085. IOP Publishing, 2017.

\bibitem{Chen2018hydrogen-assisted-MoO3-S2}
Tongxin Chen, Yingqiu Zhou, Yuewen Sheng, Xiaochen Wang, Si~Zhou, and Jamie~H
  Warner.
\newblock Hydrogen-assisted growth of large-area continuous films of mos2 on
  monolayer graphene.
\newblock {\em ACS applied materials \& interfaces}, 10(8):7304--7314, 2018.

\bibitem{Bersch2017}
Brian~M Bersch, Sarah~M Eichfeld, Yu-Chuan Lin, Kehao Zhang, Ganesh~R
  Bhimanapati, Aleksander~F Piasecki, Michael Labella, and Joshua~A Robinson.
\newblock Selective-area growth and controlled substrate coupling of transition
  metal dichalcogenides.
\newblock {\em 2D Materials}, 4(2):025083, 2017.

\bibitem{Ahn2021}
Wonsik Ahn, Hyangsook Lee, Hoijoon Kim, Mirine Leem, Heesoo Lee, Taejin Park,
  Eunha Lee, and Hyoungsub Kim.
\newblock Area-selective atomic layer deposition of mos2 using simultaneous
  deposition and etching characteristics of mocl5.
\newblock {\em physica status solidi (RRL)--Rapid Research Letters},
  15(2):2000533, 2021.

\bibitem{kim2017grains}
Hyung-Jun Kim, Hojoong Kim, Suk Yang, and Jang-Yeon Kwon.
\newblock Grains in selectively grown mos2 thin films.
\newblock {\em Small}, 13(46):1702256, 2017.

\bibitem{Villars1959}
D.~S. Villars.
\newblock A method of successive approximations for computing combustion
  equilibria on a high speed digital computer.
\newblock {\em The Journal of Physical Chemistry}, 63(4):521--525, 1959.

\bibitem{Villars1960}
D.~S. Villars.
\newblock Computation of complicated combustion equilibria on a high-speed
  digital computer.
\newblock In {\em Proceedings of the First Conference on the Kinetics
  Equilibria and Performance of High Temperature Systems, Western States
  Section of the Combustion Institute, Butterworths Scientific Publications,
  Washington DC}, volume~18, 1960.

\bibitem{Colonna2004}
G.~Colonna and A.~D'Angola.
\newblock A hierarchical approach for fast and accurate equilibrium
  calculation.
\newblock {\em Computer Physics Communications}, 163(3):177--190, 2004.

\bibitem{zeleznik1968calculation}
Frank~J Zeleznik and Sanford Gordon.
\newblock Calculation of complex chemical equilibria.
\newblock {\em Industrial \& Engineering Chemistry}, 60(6):27--57, 1968.

\bibitem{zeggeren1970computation}
F~Van~Zeggeren and SH~Storey.
\newblock The computation of chemical equilibria, cambridge university, 1970.

\bibitem{smith-missen1983}
William~R Smith and Ronald~W Missen.
\newblock {\em Chemical Reaction Equilibrium Anatysis: Theory and Algornhms}.
\newblock John Wiley, 1983.

\bibitem{Dittmer1983}
G~Dittmer and U~Niemann.
\newblock Heterogeneous reactions and chemical transport of molybdenum with
  halogens and oxygen under steady state conditions of incandescent lamps.
\newblock {\em Materials Research Bulletin}, 18(3):355--369, 1983.

\bibitem{Gurvich1990}
Lev~Veniaminovich Gurvich and I~Veyts.
\newblock {\em Thermodynamic Properties of Individual Substances: Elements and
  Compounds}, volume~2.
\newblock CRC press, 1990.

\bibitem{barin1995}
Ihsan Barin.
\newblock {\em Thermochemical Data of Pure Substances, Third Edition}.
\newblock Wiley-VCH Verlag GmbH, 1997.

\bibitem{Chase1998}
M~W Chase, editor.
\newblock {\em NIST-JANAF Themochemical Tables, Fourth Edition}.
\newblock American Chemical Society, 1998.

\bibitem{shahzad2017effects}
Rauf Shahzad, TaeWan Kim, and Sang-Woo Kang.
\newblock Effects of temperature and pressure on sulfurization of molybdenum
  nano-sheets for mos2 synthesis.
\newblock {\em Thin Solid Films}, 641:79--86, 2017.

\bibitem{yim2018wide}
Chanyoung Yim, Niall McEvoy, Sarah Riazimehr, Daniel~S Schneider, Farzan Gity,
  Scott Monaghan, Paul~K Hurley, Max~C Lemme, and Georg~S Duesberg.
\newblock Wide spectral photoresponse of layered platinum diselenide-based
  photodiodes.
\newblock {\em Nano Letters}, 18(3):1794--1800, 2018.

\bibitem{chiappe2016controlled}
Daniele Chiappe, Inge Asselberghs, Surajit Sutar, Serena Iacovo, Valeri
  Afanas'~ev, Andre Stesmans, Yashwanth Balaji, Lisanne Peters, Markus Heyne,
  Manuel Mannarino, et~al.
\newblock Controlled sulfurization process for the synthesis of large area mos2
  films and mos2/ws2 heterostructures.
\newblock {\em Advanced Materials Interfaces}, 3(4):1500635, 2016.

\bibitem{giannazzo2023atomic}
Filippo Giannazzo, Salvatore~Ethan Panasci, Emanuela Schilir{\`o}, Giuseppe
  Greco, Fabrizio Roccaforte, Gianfranco Sfuncia, Giuseppe Nicotra, Marco
  Cannas, Simonpietro Agnello, Eric Frayssinet, et~al.
\newblock Atomic resolution interface structure and vertical current injection
  in highly uniform mos2 heterojunctions with bulk gan.
\newblock {\em Applied Surface Science}, 631:157513, 2023.

\bibitem{zhang2020universal}
Tianyi Zhang, Kazunori Fujisawa, Fu~Zhang, Mingzu Liu, Michael~C Lucking,
  Rafael~Nunes Gontijo, Yu~Lei, He~Liu, Kevin Crust, Tomotaroh
  Granzier-Nakajima, et~al.
\newblock Universal in situ substitutional doping of transition metal
  dichalcogenides by liquid-phase precursor-assisted synthesis.
\newblock {\em ACS nano}, 14(4):4326--4335, 2020.

\bibitem{esposito2023role}
F~Esposito, M~Bosi, G~Attolini, F~Rossi, SE~Panasci, P~Fiorenza, F~Giannazzo,
  F~Fabbri, and L~Seravalli.
\newblock Role of density gradients in the growth dynamics of 2-dimensional
  mos2 using liquid phase molybdenum precursor in chemical vapor deposition.
\newblock {\em Applied Surface Science}, 639:158230, 2023.

\end{thebibliography}

\end{document}